\documentclass[a4paper,prd,11pt,noshowkeys,notitlepage,nofootinbib,superscriptaddress]{revtex4}
\pdfoutput=1  %necessary for JHEP
\usepackage[utf8]{inputenc}
\usepackage{ae,aecompl}
\usepackage{amsmath,amssymb,mathrsfs}
\usepackage{graphicx}
\usepackage{ccnishi}
\usepackage{dsfont}
\usepackage{slashed}
\usepackage{units}
\usepackage[normalem]{ulem}

\usepackage[usenames,dvipsnames]{xcolor} % svgnames
\usepackage{hyperref}
\hypersetup{
    colorlinks=true,        % false: boxed links; true: colored links
    linkcolor=purple,       % color of internal links, e.g., eqs.
    citecolor=JungleGreen,  % color of links to bibliography
    filecolor=magenta,      % color of file links
}

\providecommand{\cO}{\mathcal{O}}
\providecommand{\hlamb}{\hat{\lambda}}
\providecommand{\hM}{\hat{M}}

\providecommand{\hY}{\hat{Y}}
\providecommand{\hU}{\hat{U}}

\providecommand{\tMnu}{\tilde{M}_\nu}
\providecommand{\tU}{\tilde{U}}

\providecommand{\tlamb}{\tilde{\lambda}}
\providecommand{\pmns}{\mathrm{pmns}}

\providecommand{\Ueff}{U_{\rm eff}}
\providecommand{\btheta}{\bar{\theta}}

\begin{document}
%%%%%%%%%%%%%%%%%%%%%%%%%%%%%%%%%%%%%%%%%%%%%%%%%
\title{
Leptonic CP violation from the seesaw
}
\author{C.~C.~Nishi}
\thanks{E-mail: celso.nishi@ufabc.edu.br}
\affiliation{Centro de Matem\'{a}tica, Computa\c{c}\~{a}o e Cogni\c{c}\~{a}o,
Universidade Federal do ABC, 09.210-170,
Santo Andr\'{e}-SP, Brasil}
\author{J.~I.~Silva-Marcos}
\thanks{juca@cftp.tecnico.ulisboa.pt}
\affiliation{
Centro de F\'\i sica Te\'orica de Part\'\i culas -- CFTP and Dept de F\' \i sica\\
Instituto Superior T\'ecnico -- IST, Universidade de Lisboa, Av. Rovisco Pais, \\
P-1049-001 Lisboa, Portugal
}

% \date{\today}
%%%%%%%%%%%%%%%%%%%%%%%%%%%%%%%%%%%%%%%%%%%%%%%%%
\begin{abstract}
The seesaw extension of the SM explains the tiny neutrino masses and it is accompanied by many CP violating phases.
We study the case where all leptonic CP violation arises from the soft breaking in the heavy Majorana mass matrix.
Parameter counting reveals that one less parameter is needed to describe this case.
This reduction leads to restrictions on the parameter space of heavy neutrinos.
We analyze the minimal seesaw case in detail and find that mass degeneracy of heavy neutrinos is not possible for certain values of the CP phases. 
\end{abstract}
%%%%%%%%%%%%%%%%%%%%%%%%%%%%%%%%%%%%%%%%%%%%%%%%%
% \pacs{12.60.Fr, 14.80.Cp, 11.30.Qc, 02.20.-a}
%\keywords{ }
%\twocolumn
\maketitle
% \tableofcontents
%%%%%%%%%%%%%%%%%%%%%%%%%%%%%%%%%%%%%%%%%%%%%%%%%
\section{Introduction}

After the discovery that neutrinos have tiny masses and mix, one of our next major goals is to determine the existence of leptonic CP violation.
Hints for CP violating values of the Dirac CP phase $\delta$ ---analogous to the CP odd phase of the CKM mixing matrix---
is already appearing in neutrino oscillations\,\cite{t2k.2020}.
Large planned experiments such as DUNE\,\cite{dune} and Hyper-Kamiokande\,\cite{hyperK} are expected to answer this question in the future.

The existence of leptonic CP violation at low energy may be linked to the fundamental question of the origin of the matter-antimatter asymmetry of the universe which may be attributed without much effort to a leptonic CP violation at the high scale generated by the decay of the same heavy righthanded neutrinos\,\cite{leptog} that generate the tiny neutrino masses in the seesaw mechanism\,\cite{seesaw}.
It is indeed possible to have successful leptogenesis sourced by low energy CP violation, either from the Dirac or Majorana phases in the PMNS, specially in the flavored regime\,\cite{pascoli.petcov,pmns.leptogenesis}.

If we consider the usual case of the seesaw with three righthanded neutrinos (RHNs), the theory possesses 21 physical parameters, of which six are physical CP odd phases in the usual counting.
Three reside in the PMNS matrix: one Dirac CP phase and two Majorana phases.
The other three may be accessed through processes involving the RHNs at high energy or, in EFT language, in the suppressed lepton number conserving dimension six operator\,\cite{gavela:seesaw}.
In the Casas-Ibarra parametrization\,\cite{casas.ibarra}, these three phases reside in the complex orthogonal matrix $R$.\footnote{%
Note that certain special cases of CP violation are subtler than simply having complex parameters.
One example is the case of purely imaginary entries in the PMNS and purely real $R$\,\cite{pascoli.petcov}.
}
However, despite this phenomenological distinction between CP phases, theoretically, it is not clear if we can attribute a low or high scale origin to the CP violating phases.

Here we study the case where clearly leptonic CP violation ---including all the CP phases of the PMNS matrix--- has its origin at the high scale: the type I seesaw where the only source of (soft) CP breaking is the complex mass matrix of the heavy  righthanded neutrinos (RHNs).
It is straightforward to attribute this soft breaking to scalars breaking CP spontaneously.

Our study may have implications for scenarios where CP is in fact an exact symmetry of nature at high energies and CP violation at the low scale is spontaneous.
In particular, this is one of the routes to solve the strong CP problem, most notably through the Nelson-Barr setting\,\cite{nelson.barr,BBP}.
And then, leptogenesis might generate the necessary baryon asymmetry of the universe\,\cite{spontaneous.CP,reece}.
In the setting, vector-like quarks (VLQs) mix with the SM quarks to transmit the CP violation to the SM;
see Ref.\,\cite{vlq:review} for a review on singlet VLQs.
Recently, we have found that the description of these vector-like quarks of Nelson-Barr type depends on \emph{one less} parameter than the most generic case of VLQs\,\cite{nb-vlq}.
The situation is the same if leptonic CP violation is originated from a singlet charged vector-like lepton\,\cite{nb-vll}.
In all theses cases, this reduction of parameters leads to approximate patterns for the Yukawa couplings mixing the heavy fermions with the SM fermions.
We will see here that the same reduction of parameters occurs also in the case of the type I seesaw and we will investigate the possible patterns and restrictions that arise from this setting.
These resctrictions will be particularly relevant for nearly degenerate heavy Majorana neutrinos that are required for resonant leptogenesis\,\cite{resonant}.

The outline of this paper is as follows.
In Sec.\,\ref{sec:CP.M} we discuss the CP phases of the simpler case of the three active neutrinos of Majorana nature and show that the number of CP phases is parametrization dependent: one can parametrize the PMNS using three, two or just one phase.
In Sec.\,\ref{sec:typeI}, we introduce the seesaw with soft CP breaking and show that this case depends on one less parameter compared to the usual case of generic CP breaking.
We analyze the missing parameter in Sec.\,\ref{sec:missing}.
In Sec.\,\ref{sec:inv}, we study how to distinguish the case of soft CPV and generic CPV in a basis invariant manner.
Simple restrictions that follows from the missing parameter is discussed in Sec.\,\ref{sec:simple}.
The minimal case is analyzed in detail in Sec.\,\ref{sec:n=2} where we show the restrictions imposed on the parameters of the model for the case of soft CPV.
The conclusions can be found in Sec.\,\ref{sec:concl}.

%%%%%%%%%%%%%%%%%%%%%%%%%%%%%%%%%%%%%%%%%%%%%%%%%
\section{The number of CP phases for Majorana neutrinos}
\label{sec:CP.M}

Before considering the seesaw case, let us discuss the more basic case of the three active neutrinos of the SM when they are Majorana.
We will take this case as an example to show the not-well-known fact that not only the \emph{location} of 
CP phases is parametrization dependent but also its \emph{number}.
Here we are not concerned about the origin of CP violation and its origin in the UV may either be soft or hard.
The soft CPV in the seesaw will be dicussed in the rest of the paper.

Let us clarify the terminology before proceeding. Here we use the terms \emph{CP phases} and \emph{complex phases} interchangeably to mean \emph{continuous} CP \emph{odd} parameters that flip sign under the CP transformation defined in a certain CP basis where most of the Lagrangian is invariant.\footnote{We ignore the issue of radiative stability of this definition and only remark that this definition is radiatively consistent for softly broken CP.}
Then, in such a CP basis, the CP transformation acts as complex conjugation on the couplings,\footnote{See Ref.\,\cite{cp4:inert} for a caveat.} a transformation that obviously flip the sign of the complex phases and, conversely, only these phases among the parameters change by a flip.
We factor the effect of possible $i$ factors for Majorana fermions which flip sign but does not lead to CP violation.
It is important to stress that, in the same basis, one or more CP \emph{even} parameters are also needed for CP violation as a redefinition of CP might be possible interchanging their roles or CP phases may be rendered unphysical.
The well known example for the latter is the SM with degenerate quarks, with the quark mass squared differences being the CP even parameters.

Let us now turn to the PMNS.
It is well known that the location of phases in a mixing matrix connecting Dirac fermions is parametrization dependent due to rephasing freedom. For example, the unique Dirac phase $\delta$ of the CKM matrix appear in different locations in the original Kobayashi-Maskawa parametrization\,\cite{km} and in the standard parametrization\,\cite{pdg}:
\eq{
\label{V0:std}
V_0=\left(
\begin{array}{ccc}
 1 & 0 & 0 \\
 0 & c_{23} & s_{23} \\
 0 & -s_{23} & c_{23} \\
\end{array}
\right)
\mtrx{ c_{13}&0&s_{13}e^{-i\delta}\cr
    0&1&0\cr
    -s_{13}e^{i\delta}&0&c_{13}}
\mtrx{ c_{12}&s_{12}&0\cr
    -s_{12}&c_{12}&0\cr
    0&0&1}
\,,
}
using the usual shorthands $s_{ij}\equiv \sin\theta_{ij}, c_{ij}\equiv \cos\theta_{ij}$. 
We can also write it in the compact form: 
\eq{
\label{V0:std:Rij}
V_0=R_{23}(\theta_{23})U_{13}(\theta_{13},\delta)R_{12}(\theta_{12})\,.
}
These expressions define our matrices $R_{ij}$ and $U_{ij}$ where $R_{ij}(\theta)=U_{ij}(\theta,0)$.

Assuming Majorana neutrinos, the rephasing freedom from the right is lost for the PMNS matrix and two Majorana phases $\alpha,\beta$ appear in its parametrization:
\eq{
\label{pmns:std}
V_{\pmns}=V_0\mtrx{1 &&\cr &e^{i\alpha}&\cr &&e^{i\beta}}\,.
}
This parametrization employs three angles $\theta_{23},\theta_{12},\theta_{13}$, one Dirac CP phase $\delta$ and two Majorana phases $\alpha,\beta$.
One key property of this parametrization (and similar ones) is that neutrino oscillation physics does not depend on the Majorana phases $\alpha,\beta$.

We now show a parametrization that requires \emph{two} phases only:
\eq{
\label{2phases}
V_\pmns=\cO_2\diag(1,e^{i\alpha'},e^{i\beta'})R_{23}(\theta_{23}')\,,
}
where we use the convention where $V_{\pmns}$ has its first column real.
The matrix $\cO_2$ is a generic real orthogonal matrix depending on three angles.
We show in table~\ref{tab:equivalence} a numerical example of the PMNS matrix with fixed Majorana phases $\alpha,\beta$ in the standard parametrization \eqref{pmns:std} being equivalently parametrized by the two-phases parametrization \eqref{2phases}.
Together with $\theta_{23}'$ we have \emph{four} angles but only \emph{two} phases $\alpha',\beta'$.
The total number of parameters remains 6 but we traded one phase by one angle.
This parametrization is based on an orthogonal decomposition, proved in appendices \ref{ap:ortho.decomp} and \ref{ap:orthog.2}, of a generic unitary matrix $U$ in the form
\eq{
\label{ortho.param}
U=\cO_2\hat{U}\cO_1^\tp\,,
}
with $\cO_i$ being real orthogonal matrices and $\hU$ being a diagonal matrix of phases. 
With rephasing freedom from the left, one can write the parametrization \eqref{2phases}.
For a unitary matrix of generic size $n$, this kind of parametrization has potentially much less phases than the standard-like parametrization because the former has the number of phases scaling as $n$ while it is well known that the number of Dirac phases scales as $n^2$\,\cite{km}.
See appendix \ref{ap:ortho.decomp} for the exact numbers.
\begin{table}[h]
\eq{\nonumber
\begin{array}{|c|c|c|c|}
\hline
\text{Standard} & 
    \theta_{23}=42.1^\circ & \theta_{13}=8.62^\circ & \delta=230^\circ \cr & \theta_{12}=33.45^\circ & \alpha=\pi/4 & \beta=\pi/6
\cr
\hline
\text{Two phases$^*$ } &
    \btheta_{23}=20.0769^\circ & \btheta_{13}=-19.0418^\circ & \btheta_{12}=29.2266^\circ \cr & \alpha'=52.6477^\circ & \beta'=23.2009^\circ & \theta_{23}' =26.8838^\circ
\cr
\hline
\text{One phase$^{**}$} & 
    \btheta_{23}=60.0600^\circ & \btheta_{13}=-68.5049^\circ & \btheta_{12}=13.6452^\circ \cr & \beta''=-105.12^\circ & \theta_{23}''=83.6798^\circ & \theta_{12}'' =54.1295^\circ
\cr
\hline
\end{array}
}
\caption{\label{tab:equivalence}%
An example Majorana PMNS matrix parametrized in the standard parametrization, in the two-phases parametrization and in the one-phase parametrization, respectively; cf.\,\eqref{V0:std}, \eqref{2phases} and \eqref{1phase}.
The parameters in the first row except $\alpha,\beta$ are taken from the global fit in Ref.\,\cite{nufit} with SK data.
$^*$ Rephasing convention from the left: $V_{11}>0, V_{31}>0$ and $V_{21}<0$.
We use $\cO_2=R_{12}R_{13}R_{23}$ for the explicit parametrization of $\cO_2$ in \eqref{2phases} and \eqref{1phase}, with the barred angles. $^{**}$ Rephasing from the left and sign flips from the right should be used to check equivalence. 
}
\end{table}

Specifically for matrices of size three, we can minimize even further and find a parametrization depending on \emph{one} phase\,\cite{juca}:
\eq{
\label{1phase}
V_\pmns=\cO_2\diag(1,1,e^{i\beta''})R_{23}(\theta_{23}'')R_{12}(\theta_{12}'')\,,
}
where again $\cO_2$ is a generic real orthogonal matrix with three angles.
Thus we have \emph{five} angles and \emph{one} phase.
This is valid for matrices with rephasing freedom from the left.
See table~\ref{tab:equivalence} for the same numerical example being parametrized with one phase,
assuming additional sign flip freedom from the right.

Therefore, generically, the number of CP phases is parametrization dependent and we should supply the parametrization when we count CP phases.\footnote{%
In contrast, the number of \emph{algebraically independent} CP odd \emph{polynomial flavor invariants} is well defined\,\cite{invariants:others}.
}
The total number of parameters on the other hand is a robust quantity independent of the chosen parametrization.
We should also emphasize that the \emph{presence} or \emph{absence} of CP violation is also a physical property independent of parametrization. It can be formulated in a basis invariant manner if necessary.
Then in the presence of CP violation, it follows that \emph{at least} one nonzero physical phase should be present in any parametrization.

%%%%%%%%%%%%%%%%%%%%%%%%%%%%%%%%%%%%%%%%%%%%%%%%%
\section{Type I seesaw with soft CP breaking}
\label{sec:typeI}

We now treat the type I seesaw extension of the SM with $n_R$ righthanded neutrinos (RHNs) $N_{iR}$ described by the lagrangian:
\eq{
\label{lag:seesaw}
-{\cal L} =
Y_{\alpha\beta} \bar \ell_\alpha H e_{\alpha R}
+ \lambda_{i\alpha} \bar N_{iR} \tilde H^\dagger \ell_\alpha
+\ums{2}\bar N_{iR} (M_R)_{ij}N_{jR}^c 
+h.c.
}
Sometimes we will adopt the basis where $Y$ is diagonal, denoted by $\hY=\diag(y_\alpha)$, $\alpha=e,\mu,\tau$.
Light neutrino masses are generated by $N_R$ exchange leading to the seesaw relation
\eq{
\label{Mnu:ss}
M_\nu=-v^2\lambda^\tp M_R^{-1}\lambda
\,,
}
where $v=\aver{H^0}=174\,\unit{GeV}$.

We want to compare two cases: (i) the generic (usual) case where CP is explicitly broken ($Y,\lambda$ and $M_R$ complex) and (ii) the case where CP symmetry is only softly broken in the Majorana mass term $M_R$ (the rest of the couplings are real). This soft breaking can be obviously spontaneous, induced by appropriate singlet scalars getting CP breaking vevs.
We consider the theory at the seesaw scale or below where these scalars are integrated out and the prime   feature is the soft CP breaking.
To distinguish the cases (i) and (ii), we will denote them as \emph{generic} and \emph{soft} cases of CP violation (CPV).

The first important observation is that the soft CPV case depends on \emph{one less} parameter than the generic CPV case. This feature is common to many situations involving soft (spontaneous) CP breaking\,\cite{nb-vlq,nb-vll}.
We show\,\footnote{%
Note that the distinction between generic and soft CPV starts at $n_R=2$ because for $n_R=1$ all phases can be eliminated and CP is conserved.
In any case, we need $n_R\ge 2$ to generate at least two massive active neutrinos.
}
in table\,\ref{tab.1}
the number of coupling parameters of the seesaw lagrangian \eqref{lag:seesaw} for the generic and soft CPV cases as well as for the CP conserving case (all couplings real).
Due to the discussion of Sec.\,\ref{sec:CP.M} about the difficulty of counting CP odd parameters (phases), we refrain from giving this number.
The usual counting of moduli and phases for the generic case can be obtained from the last column for the former and from the difference between the second column and the last for the second. These numbers for generic $n_R$ can be seen, e.g., in table~1 of Ref.\,\cite{gavela:seesaw}.
In the next section we show explicitly for $n_R=3$ the parameter that gets removed when we go from the generic to the soft CPV case in a given parametrization.
There we also briefly describe one possible basis where the counting can be performed.
\begin{table}[h]
\[
\begin{array}{|c|c|c|c|}
\hline
\text{Type I seesaw} & \text{Generic CPV} & \text{Soft CPV} & \text{CP cons.}
\cr
\text{$n_R$ RHNs}    & \text{\# param.} & \text{\# param.} &\text{\# param.} \cr
\hline
2 & 14  & 13 & 11 \cr
3 & 21  & 20 & 15 \cr
4 & 28  & 27 & 19 \cr
\hline
\end{array}
\]
\caption{\label{tab.1}%
Number of coupling parameters of the seesaw lagrangian \eqref{lag:seesaw} in three cases: generic CPV, soft CPV and CP conservation. 
The usual number of CP phases in the generic CPV case can be obtained from the difference of the second column and the last one.
}
\end{table}

%%%%%%%%%%%%%%%%%%%%%%%%%%%%%%%%%%%%%%%%%%%%%%%%%
\section{The missing phase}
\label{sec:missing}

Let us identify how the Lagrangian \eqref{lag:seesaw} with couplings $Y_{\alpha\beta},\lambda_{i\alpha},(M_R)_{ij}$ loses one parameter when we go from the generic ($Y,\lambda,M_R$ complex) to the soft (only $M_R$ complex) CPV case. We will see that this missing parameter may be counted as a phase.
In the process we will also show briefly how the counting in table~\ref{tab.1} is made.

We first treat the soft CPV case where $Y,\lambda$ are real. Orthogonal rotations on $\ell_\alpha, e_{\alpha R}$ and $N_{iR}$ allow us to choose a basis where 
\eq{
\label{basis:lambda}
\lambda=\hlamb\,,\quad
Y=O_e\hY\,,
\qquad[\text{diagonal-$\lambda$ basis (soft CPV)}]
}
with $\hlamb$ containing the singular values in the main diagonal although it may not be a square matrix as $\lambda\sim n_R\times 3$ with $n_R\ge 2$.
The matrix $O_e$ is a $3\times 3$ real orthogonal matrix.
Further rephasing is not allowed as it would disrupt real $Y,\lambda$.
The only possible basis change left is global rephasing along total lepton number which removes one parameter in $M_R$.

The counting of parameters in this basis leads to the third column of table~\ref{tab.1}.
For example, for $n_R=3$, we have
\eq{
\label{count:n=3:scpv}
Y\sim 3+3\,,\quad
\lambda\sim 3\,,\quad
M_R\sim 3(3+1)-1=11\,.
}

To connect to a more physical basis, we can diagonalize 
\eq{
M_R\to U_R^\dag M_RU_R^*=\hM_R\,,
}
where $U_R\in SU(n_R)$ when we choose $\det M_R$ real positive.
We also transfer $O_e$ from $Y$ to $\lambda$ in \eqref{basis:lambda} so that $Y\to \hY$ becomes diagonal.
Compared to the basis \eqref{basis:lambda}, we denote the Yukawa in the resulting basis as
\eq{
\label{lambdaR:soft}
\lambda_R\equiv U_R^\dag \hlamb O_e\,
\qquad[M_R=\hM_R\,,~Y=\hY].
}
Then it obeys the reality condition
\eq{
\label{real.cond}
(\lambda^\dag_R \lambda_R)_{\alpha\beta}\sim\text{real}
\qquad
\text{(Soft CPV)}\,.
}
This relation is invariant by unitary rotations on $N_{iR}$.
At the same time, CP violation still manifests in complex $\lambda_R\lambda_R^\dag$ or complex $U_R$ in \eqref{lambdaR:soft}.

We can now review the generic CPV case which is well known.
The analysis is simple in the basis where $M_R$ is diagonal and $\lambda_R$ is complex generic, except for rephasing from the right, which is now allowed.

However, to compare with the soft CPV case, we can adopt the basis analogous to \eqref{basis:lambda} where
\eq{
\label{basis:lambda:hard}
\lambda=\hlamb\,,\quad
Y=U_e\hY\,,
\qquad[\text{diagonal-$\lambda$ basis (generic CPV)}]
}
where $U_e$ is now unitary.
This is achieved by unitary rotations on $\ell_\alpha, e_{\alpha R}$ and $N_{iR}$.
Rephasing on $\ell_\alpha,e_{\alpha R}$ induces rephasing on both sides of $U_e$ and it contains only \emph{one} CKM-like phase.
This is the additional phase that is missing in the soft CPV case.
The matrix $M_R$ is still generic with one less parameter along lepton number. 

Parameter counting easily leads to the second column of table~\ref{tab.1}.
For example, for $n_R=3$, we can count analogously to \eqref{count:n=3:scpv}:
\eq{
Y\sim 3+4,\quad
\lambda\sim 3\,,\quad
M_R\sim 3(3+1)-1\,.
}
In the basis of diagonal $Y$ and $M_R$, eq.\,\eqref{lambdaR:soft} is modified to
\eq{
\label{lambdaR:hard}
\lambda_R\equiv U_R^\dag \hlamb U_e\,
\qquad[M_R=\hM_R\,,~Y=\hY].
}
The matrix $U_e$ contains one additional phase compared to $O_e$ in \eqref{lambdaR:soft}.

The reality condition \eqref{real.cond} does not apply and we have, modulo rephasing, 
\eq{
\label{not.real.cond}
(\lambda_R^\dag \lambda_R)_{\alpha\beta}\sim\text{complex}
\qquad
\text{(Generic CPV)}\,.
}
Since $\lambda^\dag_R\lambda_R$ is $3\times 3$, it contains only one physical phase after rephasing, the same number of phases in $U_e$.

%%%%%%%%%%%%%%%%%%%%%%%%%%%%%%%%%%%%%%%%%%%%%%%%%
\section{Invariant conditions for soft CPV}
\label{sec:inv}

Once the presence of some CP violation is detected, the \emph{distinction} between the soft CPV case, cf.\,\eqref{real.cond}, and the generic CPV, cf.\,\eqref{not.real.cond}, resides in only one phase and the condition can be formulated in a \emph{rephasing} invariant way: CPV in the seesaw is soft if and only if
\eq{
\label{rephasing:lambda}
\im[(\lambda^\dag\lambda)_{e\mu}(\lambda^\dag\lambda)_{\mu\tau}(\lambda^\dag\lambda)_{\tau e}]
=0\,,
\quad\text{(rephasing inv.)},
}
in the basis where the charged leptons Yukawas $Y$ are diagonal; $M_R$ can be in any basis.
In a general weak basis, for the couplings in the Lagrangian \eqref{lag:seesaw}, we could  define 
\eq{
H_e\equiv YY^\dag, \quad H_\lambda\equiv\lambda^\dag\lambda\,,
}
and formulate the equivalent \emph{basis} invariant condition: CPV in the seesaw is soft if and only if
\eq{
\label{inv:HH}
\im\Tr\Big[[H_e,H_\lambda]^3\Big]=0
\quad\text{(basis inv.)}.
}
This is similar to the quark sector\,\cite{jarlskog:comm}.

One can also formulate the rephasing invariant condition \eqref{rephasing:lambda} in terms of the matrix $U_e$ that diagonalizes
\eq{
U_e^\dag\lambda^\dag \lambda U_e=\hlamb^\dag \hlamb\sim \text{diagonal}\,.
}
This is the same matrix in \eqref{basis:lambda:hard}, modulo rephasing of the eigenvectors which is allowed here.
Then, CPV in the seesaw is soft if and only if
\eq{
\label{cond:J}
J(U_e)=0\,,
}
where $J$ is one of the Jarlskog invariants that can be constructed with a $3\times 3$ unitary matrix 
familiar from the CKM matrix\,\cite{jarlskog:comm}.
A simple choice is $J(U)=\im(U_{11}U_{22}U_{12}^*U_{21}^*)$.

Of course, to distinguish soft CPV from CP conservation, one needs CP violating couplings somewhere.
Once the condition for soft CPV is satisfied in \eqref{rephasing:lambda} or \eqref{inv:HH} or \eqref{cond:J}, one can go to the basis in \eqref{basis:lambda} and explicitly check if $M_R$ is complex even with the removal of a global phase.
Since it involves many phases, this procedure is easier than checking basis invariant conditions\,\cite{dreiner,branco.morozumi,invariants:others}.\footnote{%
In special cases, such as mass degeneracy, more subtleties are involved.\,\cite{yu.zhou}.
}

%%%%%%%%%%%%%%%%%%%%%%%%%%%%%%%%%%%%%%%%%%%%%%%%%
\section{No restriction on the PMNS and simple restrictions}
\label{sec:simple}

In Sec.\,\ref{sec:missing}, we saw that the diagonal-$\lambda$ basis was appropriate to compare the soft CPV case \eqref{basis:lambda} with the generic CPV case \eqref{basis:lambda:hard}.
The difference consisted in one phase contained in $U_e$ but absent in $O_e$ for the soft CPV case.
This unitary (or real orthogonal) matrix will enter in the PMNS matrix as
\eq{
\label{V:lambda-diag}
\text{soft CPV:~~~}
V=O_e^\dag U_\nu\,;\qquad
\text{generic CPV:~~~}
V=U_e^\dag U_\nu\,.
}
The matrix $U_\nu$ is the matrix that diagonalizes the light neutrino mass matrix:
\eq{
\label{Mnu.diag}
U_\nu^\tp M_\nu U_\nu=\hM_\nu
=\diag(m_1,m_2,m_3)
\,.
}
Note that the light neutrino mass matrix $M_\nu$ is still given by the seesaw relation \eqref{Mnu:ss} but it is not in the usual basis of diagonal charged lepton Yukawa $Y$.

Since $M_R$ is generic in the diagonal-$\lambda$ basis, the light neutrino mass matrix $M_\nu$ can be generic in the seesaw relation and so can be $U_\nu$.
Therefore, it is clear that we have enough freedom to obtain any PMNS matrix in \eqref{V:lambda-diag} even in the soft-CPV case.
So there is no restriction on the PMNS.

However, once we \emph{require} that $V$ be the physical PMNS matrix with known mixing angles, \emph{restrictions} will appear on the matrix $U_\nu$ for \emph{the soft CPV case}.
It is clear from \eqref{V:lambda-diag} that if $V$ contains a nontrivial Dirac CP phase $\delta$, the matrix $U_\nu$ cannot be any matrix.
For example, the identity matrix, a real matrix or a diagonal matrix of phases are all excluded in that case.
Ultimately, the CP violation of the PMNS has to be provided by $U_\nu$.

In turn, once $U_\nu$ is resctricted to account for the physical PMNS, this restriction will lead to a restriction on the heavy neutrino mass matrix $M_R$ through the seesaw relation 
\eq{
\label{seesaw:Unu}
U_\nu^*\hM_\nu U_\nu^\dag
=-v^2\hlamb^\tp M_R^{-1}\hlamb\,.
}

We will study in detail the case of two RHNs in Sec.\,\ref{sec:n=2}.
Let us briefly discuss here the case of soft CPV for three RHNs: $n_R=3$.
In this case, using the PMNS matrix $V$ as input,\footnote{%
The matrix $V$ should include two more phases from the left compared to the usual parametrization in \eqref{pmns:std}:
$V=\diag(1,e^{-i\alpha'},e^{-i\beta'})
V_0\diag(1,e^{i\alpha},e^{i\beta})$.
The phases $\alpha',\beta'$ are necessary because the righthand side of \eqref{V:lambda-diag} (soft CPV) does not automatically lead to the correct rephasing convention corresponding to \eqref{pmns:std}.
These phases are not observable from low energy physics.
They only affect $U_\nu$ in \eqref{Unu:n=3:soft} which in turn affect $M_R$ in \eqref{seesaw:Unu}.
} 
we can invert the soft CPV relation \eqref{V:lambda-diag} as
\eq{
\label{Unu:n=3:soft}
U_\nu=O_e V\,,
}
where $O_e$ is a generic real orthogonal matrix in \eqref{basis:lambda} with three angles.
Then the matrix $U_\nu$ can be determined from $O_e$ and $V$.
It is clear that if $V$ is complex and contains a nontrivial Dirac CP phase, a real $O_e$ cannot fully compensate $V$ and $U_\nu$ cannot reach the identity.

In contrast, in the generic CPV case, we should replace $O_e$ by a unitary matrix $U_e$.
Then it can fully cancel $V$ and lead to any possible $U_\nu$.
So in the generic CPV case, there is no restriction on $U_\nu$ and hence no restriction on $M_R$.

Let us end this section by emphasizing that it is not possible to use the Casas-Ibarra parametrization in the soft CPV case.
In other words, we cannot simultaneously use the
physical information of the PMNS matrix, light neutrino masses and heavy neutrino masses as input.

The reason for the difficulty is that new constraints arise.
We illustrate this point again for $n_R=3$.
In this case, the Casas-Ibarra parametrization\,\cite{casas.ibarra} is
\eq{
\label{casas.ibarra}
v\lambda_R=i\hM_R^{1/2}R\hM_\nu^{1/2}V^\dag\,,
}
where $R$ is a complex orthogonal matrix parametrizing the additional freedom besides the heavy neutrino masses in $\hM_R$, the light neutrino masses in $\hM_\nu$ and the PMNS matrix in $V$.
The Yukawa coupling $\lambda_R$ is defined in the basis where $M_R$ and the charged lepton Yukawa $Y$ are diagonal.

However, for the soft CP case, $\lambda_R$ should obey the reality condition \eqref{real.cond} which in the Casas-Ibarra parametrization reads
\eq{
v^2\lambda_R^\dag\lambda_R=
V\hM_\nu^{1/2}R^\dag \hM_R R\hM_\nu^{1/2} V^\dag\sim \text{real}\,,
}
modulo rephasing.
This means that the imaginary part of $R$ needs to somehow cancel the imaginary part of the PMNS $V$. This constraint is hard to implement.

As a final discussion of this section, let us consider the simple consequences on the CP asymmetry responsible for leptogenesis.
Let us write the flavored CP asymmetry in the decay $N_1\to \ell_\alpha\nu_\alpha$\,\cite{cp.asym}:
\eqali{
\label{cpasym:abc}
\eps_{1\alpha}&=
-\sum_{k\neq 1}
\frac{\im(\lambda^\dag_{\alpha 1}(\lambda\lambda^\dag)_{1k}\lambda_{k\alpha})}
{8\pi(\lambda\lambda^\dag)_{11}}
\frac{1}{1-x_k}
\cr
&
\quad -\ \sum_{k\neq 1}
\frac{\im(\lambda^\dag_{\alpha 1}(\lambda\lambda^\dag)_{k1}\lambda_{k\alpha})}
{8\pi(\lambda\lambda^\dag)_{11}}
\sqrt{x_k}\left[
\frac{1}{1-x_k}+1-(1+x_k)\ln\Big(1+\frac{1}{x_k}\Big)
\right]
\,,
}
where $x_k=M_k^2/M_1^2$.
In these expressions $\lambda$ means $\lambda_R$.
The unflavored CP asymmetry is obtained from \eqref{cpasym:abc} by summing in $\alpha$.
Then the first line in \eqref{cpasym:abc}, which conserves total lepton number, vanishes.
The second line has the imaginary part dependent only on $\im[(\lambda_R\lambda_R^\dag)_{k1}]^2$.
Then, given the structure $\lambda_R=U_R^\dag\hlamb U_e$ in the generic CPV case, cf.\,\eqref{lambdaR:hard},
or $\lambda_R=U_R^\dag\hlamb O_e$ in the soft CPV case, the unflavored CP asymmetry does not depend on the additional phase present in $U_e$ of the generic CPV case.
The difference between generic CPV and soft CPV would appear only through possible restrictions on the spectrum of the RHNs and on $U_R$ due to the restriction of $U_\nu$. We show this restrictions for the minimal case in Sec.\,\ref{sec:n=2}.

%%%%%%%%%%%%%%%%%%%%%%%%%%%%%%%%%%%%%%%%%%%%%%%%%
\section{The minimal seesaw case: $n_R=2$}
\label{sec:n=2}

Here we study in detail the soft CPV case for the minimal case of two RHNs\,\cite{min.seesaw} where the reduced number of parameters will allow us to be quantitative and often we will be able to extract analytical expressions. 
Comparison to the generic CPV case will be made on appropriate places.

\subsection{Parametrization in the diagonal-$\lambda$ basis}

Let us device an explicit parametrization for $n_R=2$ starting from the diagonal-$\lambda$ basis in \eqref{basis:lambda}.
Considering that for NO (IO) the massless neutrino is $\nu_1$ ($\nu_3$), we choose the first (third) column of $\lambda$ to be zero:
\eq{
\label{lambda-diag:n=2}
\text{NO:}\quad
\hlamb=\mtrx{0&\lambda_1&0\cr 0&0&\lambda_2}
=\Bigg(\begin{array}{c}0\cr 0\end{array}\Bigg|~\tlamb~\Bigg)
\,,
\quad
\text{IO:}\quad
\hlamb=\mtrx{\lambda_1&0&0\cr 0&\lambda_2&0}
=\Bigg(~\tlamb~\Bigg|\begin{array}{c}0\cr 0\end{array}\Bigg)
\,,
}
where $\tlamb=\diag(\lambda_1,\lambda_2)$ is the nonsingular square block.
The charged lepton Yukawa has the form \eqref{basis:lambda} for the soft CPV case and \eqref{basis:lambda:hard} for the generic CPV case. The heavy mass matrix $M_R$ is generic.

With these choices for $\lambda$, the light neutrino mass matrix from the seesaw formula \eqref{Mnu:ss} yields the block diagonal form:
\eq{
\label{n=2:Mnu}
\text{NO:}\quad
M_\nu=\mtrx{0&\cr &\tMnu}\,,
\quad
\text{IO:}\quad
M_\nu=\mtrx{\tMnu &\cr &0}\,,
}
where, for both cases, the nontrivial $2\times 2$ block is
\eq{
\label{n=2:Mtil}
\tMnu=-v^2\tlamb M_R^{-1}\tlamb\,.
}

The matrices in \eqref{n=2:Mnu} are evidently diagonalized by a block diagonal unitary matrix
\eq{
\label{n=2:Unu}
\text{NO:}\quad
U_\nu=i\mtrx{1&\cr &\tU}\,,
\quad
\text{IO:}\quad
U_\nu=i\mtrx{\tU &\cr &1}\,.
}
We have included the factor $i$ for convenience and conventionally keep $\tU$ within $SU(2)$.

The PMNS matrix is given by the mismatch \eqref{V:lambda-diag} of $U_\nu$ and the diagonalizing matrix of the charged lepton Yukawa $Y$. 
Given the block structure in \eqref{n=2:Unu}, discarding the $i$ factor,
it follows that the PMNS matrix for the CPV case has the first (third) column real for NO (IO).
For a proper comparison, we should use the same phase convention for the generic CPV case.

To extract the consequences on $M_R$, we can invert \eqref{n=2:Mtil}.
If we take NO, it reads
\eq{
\label{n=2:MR:Utilde}
M_R=v^2\tlamb \tU\diag(m_2^{-1},m_3^{-1})\tU^\tp \tlamb\,.
}
For IO, we just replace the light neutrino masses $m_2\to m_1,m_3\to m_2$.
For further analysis, it is more convenient to normalize with respect to 
\eq{
\label{def:M0}
M_0\equiv |\det M_R|^{1/2}=\sqrt{M_1M_2}=v^2\sqrt{\frac{\lambda_1^2\lambda_2^2}{m_2m_3}}\,,
}
where $M_1,M_2$ are the heavy neutrino masses.
Then we can parametrize
\eq{
\label{MR:norm}
\frac{M_R}{M_0}
=\diag(\kappa,\kappa^{-1})\tU\diag(\rho,\rho^{-1})\tU^\tp \diag(\kappa,\kappa^{-1})
\,,
}
where $\rho_{\rm NO}\equiv \sqrt{m_3/m_2}=(\Delta m^2_{\rm atm}/\Delta m^2_{\rm sol})^{1/4}\approx 2.41$ for NO.
For IO, we should use $\rho_{\rm IO}\equiv\sqrt{m_2/m_1}\approx 1.0076$.
The other variable is $\kappa=\sqrt{\lambda_1/\lambda_2}$.

%%%%%
\subsection{Parametrization of $V$}

Given the special feature of the PMNS matrix having the first (third) column real for NO (IO), it is more 
convenient to adopt a parametrization that makes this feature explicit.

We will adopt
\eqali{
\label{param:NO-IO}
V\big|_{\rm NO}&=R_{23}(-\theta_1)R_{12}(-\theta_2)
\mtrx{1&\cr & W(\alpha,\theta_3,\beta)}\,,
\cr
V\big|_{\rm IO}&=R_{13}(\theta_1)R_{23}(\theta_2)
\mtrx{W(\alpha,\theta_3,\beta) &\cr &1}\,,
}
where $R_{ij}$ are defined in \eqref{V0:std} and  \eqref{V0:std:Rij}, while $W$ is a $2\times 2$ representation of $SU(2)$ defined by
\eq{
\label{def:W}
W(\alpha,\theta,\beta)\equiv e^{-i\sigma_3\alpha/2}e^{-i\sigma_2\theta}e^{-i\sigma_3\beta/2}
    =\mtrx{c_\theta e^{-i(\alpha+\beta)/2} & -s_\theta e^{-i(\alpha-\beta)/2}
    \cr 
    s_\theta e^{+i(\alpha-\beta)/2} & c_\theta e^{+i(\alpha+\beta)/2}
    }
\,,
}
using the notation with the Pauli matrices $\sigma_i$.
The parametrization assumes $\theta_i\in [0,\pi/2]$, $\alpha\in [0,2\pi]$ and $\beta\in [0,\pi]$.
We can note that $\beta$ in \eqref{param:NO-IO} is the single Majorana phase of the minimal case and it is not currently constrained.
An equivalent form for the last matrix in \eqref{def:W} is $\diag(1,e^{i\beta})$, in which case it is clear that $\beta$ goes with $\nu_3$ for NO and with $\nu_2$ for IO. 

\begin{figure}[h]
\includegraphics[scale=.55]{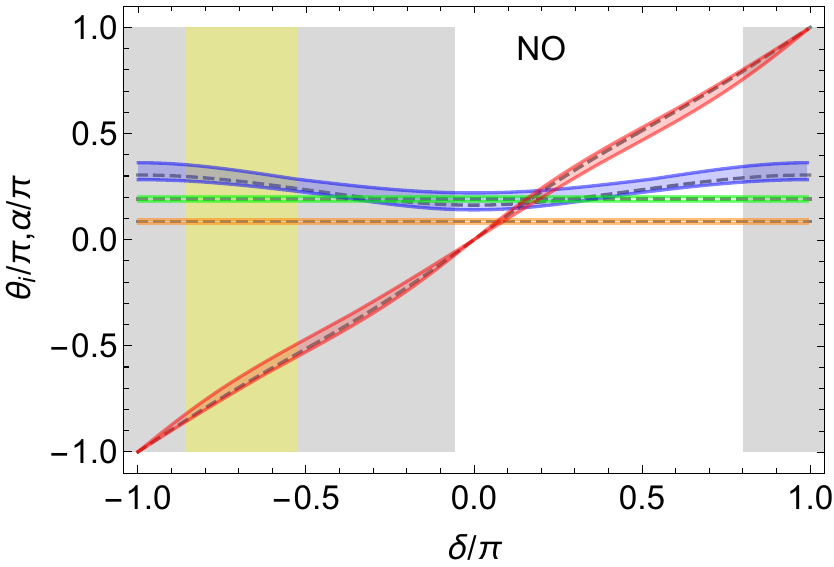}
\includegraphics[scale=.55]{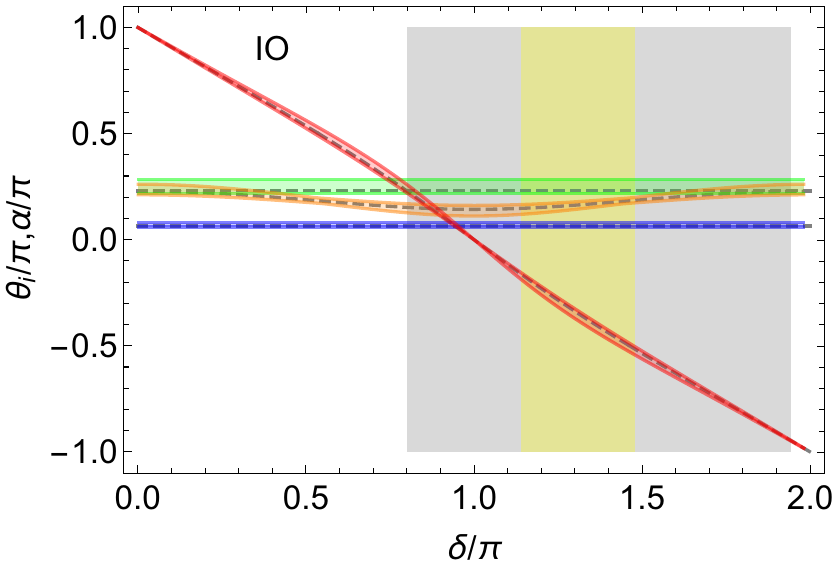}
\caption{\label{fig:theta-delta}
Angles $\theta_1$ (blue), $\theta_2$ (green), $\theta_3$ (orange) and phase $\alpha$ (red) in
\eqref{param:NO-IO} as a function $\delta$.
The bands denote $3\sigma$ variation of the mixing angles while the dashed curves denote the use of best-fit values. The gray (yellow) region denotes the current $3\sigma$ ($1\sigma$) region for $\delta$.
}
\end{figure}

The parametrization for NO in \eqref{param:NO-IO} has indeed the first column real and additionally positive entries when $\theta_1,\theta_2$ lie in the first quadrant.
Similarly, the parametrization for IO has the third column real and positive when $\theta_1,\theta_2$ lie in the first quadrant.
The NO parametrization has the additional advantage that $\theta_3$, the angle that is accompanied by complex phases, is ensured to be small as $\theta_3=0$ when $V_{13}=0$.
Figure~\ref{fig:theta-delta} shows how the various angles and phase in \eqref{param:NO-IO} depends on the Dirac phase $\delta$ of the standard parametrization.
We can see that the angles $\theta_i$ depend only mildly on $\delta$.
The extraction of these parameters is similar to the case of standard-like parametrizations\,\cite{denton}.

\begin{figure}[h]
\includegraphics[scale=.55]{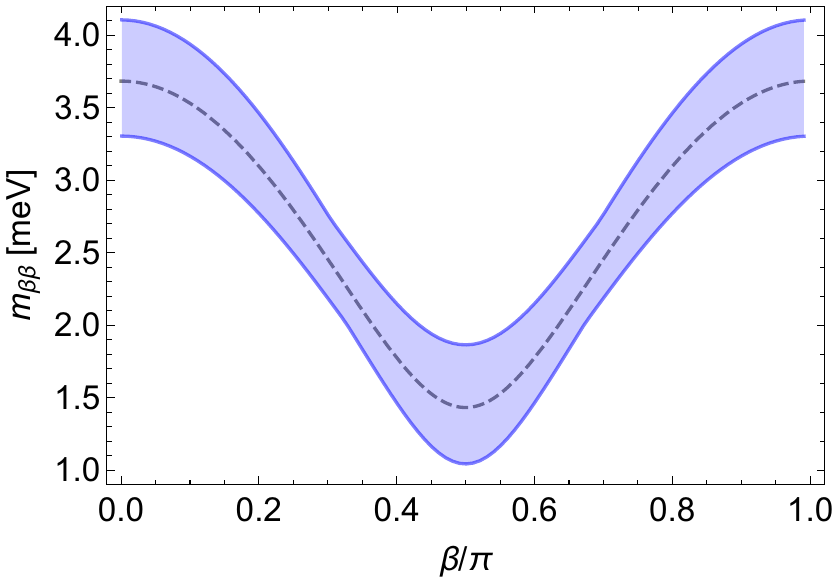}
\includegraphics[scale=.55]{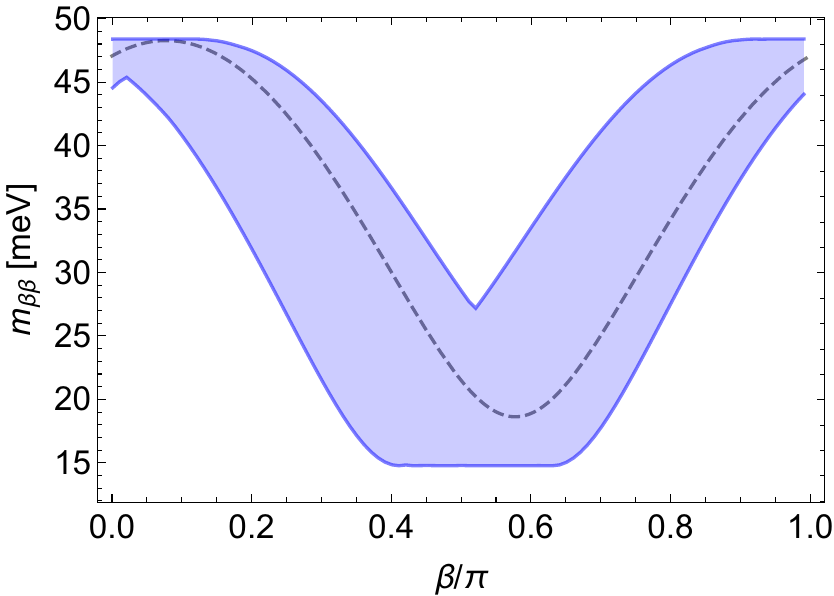}
\caption{\label{fig:beta-mbetabeta}
Effective mass for neutrinoless double beta decay as a function of the Majorana phase $\beta$ for NO (left) and IO (right).
The bands denote the variation of the PMNS parameters within $3\sigma$ while the dashed lines denote the use of best-fit values. 
}
\end{figure}

In Fig.\,\ref{fig:beta-mbetabeta} we also show the effective mass $m_{\beta\beta}$ of neutrinoless double beta decay as a function of the single Majorana phase $\beta$ as defined in the parametrizations of \eqref{param:NO-IO}.
The bands denote the $3\sigma$ variation of parameters of the PMNS matrix while the dashed lines denote the use of best-fit values. 
In the NO parametrization, $m_{\beta\beta}$ does not depend on $\delta$ while, in the IO parametrization, the dependence on $\delta$ is mild.
Current and future neutrinoless double beta decay experiments will soon start to probe the IO region around 50\,\unit{meV}; see a review in \cite{nuless:review}.

\subsection{Restrictions on $\tU$}

As shown in Sec.\,\ref{sec:simple}, the unitary matrix $U_\nu$ that diagonalizes the light neutrino mass matrix $M_\nu$ in the diagonal-$\lambda$ basis is restricted in the soft CP case once we take into account the physical PMNS, and provided that the latter violates CP.
For the minimal case, the nontrivial block of $U_\nu$ is $\tU$, cf.~\eqref{n=2:Unu}.
Here, we will show the explicit restrictions on $\tU$.
For the generic CPV case, $\tU$ is free.

Focusing on NO and discarding the global $i$ factor in \eqref{n=2:Unu}, the PMNS matrix is given by
\eq{
\label{V:n=2}
V=O_e^\tp \mtrx{1&\cr &\tU}\,.
}
It should be clear that the first column of $V$ coincides with the first column of $O_e^\tp$, hence it is real. For IO, the same occurs for the third column.
As a general restriction, $\tU$ cannot be the identity if we want a complex $V$.
In contrast, for the generic CPV case we understand that $\tU$ is free in the following manner: the real $O_e^\tp$ becomes the unitary $U_e^\dag$ and this matrix can cancel any $\tU$ and can still give the correct $V$.

Given the form of $V$ in \eqref{param:NO-IO}, it is convenient to adopt the following parametrizations for 
$O_e$:
\eqali{
\label{param:n=2:Oe}
O_e^\tp\Big|_{\rm NO}&=R_{23}(-\theta_1)R_{12}(-\theta_2)
R_{23}(-\theta_3')\,,
\cr
O_e^\tp\Big|_{\rm IO}&=R_{13}(\theta_1)R_{23}(\theta_2)
R_{12}(-\theta_3')\,.
}
We can adopt $\theta_1,\theta_2\in [0,\pi/2]$ to keep the first (third) column positive for NO (IO).
We can also take $\theta_3'\in [0,\pi]$ as some sign flips can be absorbed by Majorana fields. 

With the previous parametrization of $O_e^\tp$, we can identify in $V$ the structure
\eqali{
\label{pmns:n=2}
V\big|_{\rm NO}&=R_{23}(-\theta_1)R_{12}(-\theta_2)\mtrx{1&\cr & \Ueff}\,,
\cr
V\big|_{\rm IO}&=R_{13}(\theta_1)R_{23}(\theta_2)\mtrx{\Ueff & \cr &1}\,,
}
where the effective subblock is 
\eq{
\label{Ueff:tildeU}
\Ueff= e^{-i\sigma_2\theta_3'}\tU\,.
}

Comparing to the parametrization of $V$ in
\eqref{param:NO-IO}, we conclude that the angles $\theta_1,\theta_2$ are determined from the current knowledge of the PMNS, except for $\delta$. They are given in Fig.\,\ref{fig:theta-delta}.
The effective subblock $\Ueff$ is also mostly determined.
Using the parametrization \eqref{def:W} of $SU(2)$, 
we have
\eq{
\label{Ueff:n=2}
\Ueff=W(\alpha,\theta_3,\beta)\,,
}
where $\theta_3$ and $\alpha$ are also given in Fig.\,\ref{fig:theta-delta} as a function of $\delta$.
Therefore, inverting \eqref{Ueff:tildeU}, $\tU$ is given by
\eq{
\label{tU:W}
\tU=e^{+i\sigma_2\theta_3'}W(\alpha,\theta_3,\beta)\,,
}
where $\theta_3'$ and $\beta$ are free.

If we parametrize $\tU$ as
\eq{
\label{tU:param}
\tU=W(\alpha_\nu,\theta_\nu,\beta_\nu)\,,
}
then $\alpha_\nu,\theta_\nu$ will be restricted by the fact that $\alpha,\theta_3$ are functions of $\delta$ in \eqref{tU:W}.
The latter leads to the relation
\eq{
\label{Utilde:OPO}
e^{-i\sigma_3\alpha_\nu/2}e^{-i\sigma_2\theta_\nu}e^{-i\sigma_3(\beta_\nu-\beta)/2}
=e^{+i\sigma_2\theta_3'}
e^{-i\sigma_3\alpha/2}e^{-i\sigma_2\theta_3}
\,,
}
or
\eq{
\label{Utilde:POP}
e^{-i\sigma_2\theta_3'}e^{-i\sigma_3\alpha_\nu/2}e^{-i\sigma_2\theta_\nu}
=
e^{-i\sigma_3\alpha/2}e^{-i\sigma_2\theta_3}e^{-i\sigma_3(\beta-\beta_\nu)/2}
\,.
}
In these equations, both sides are generic parametrizations of $SU(2)$ and then one side can be written as a function of the other side.
One parametrization is the usual phase-angle-phase parametrization in \eqref{def:W} and the other makes use of the orthogonal decomposition of 
appendix~\ref{ap:ortho.decomp}.
For definiteness, we will use \eqref{Utilde:OPO} for most of the calculations.

\begin{figure}[h]
\includegraphics[scale=.45]{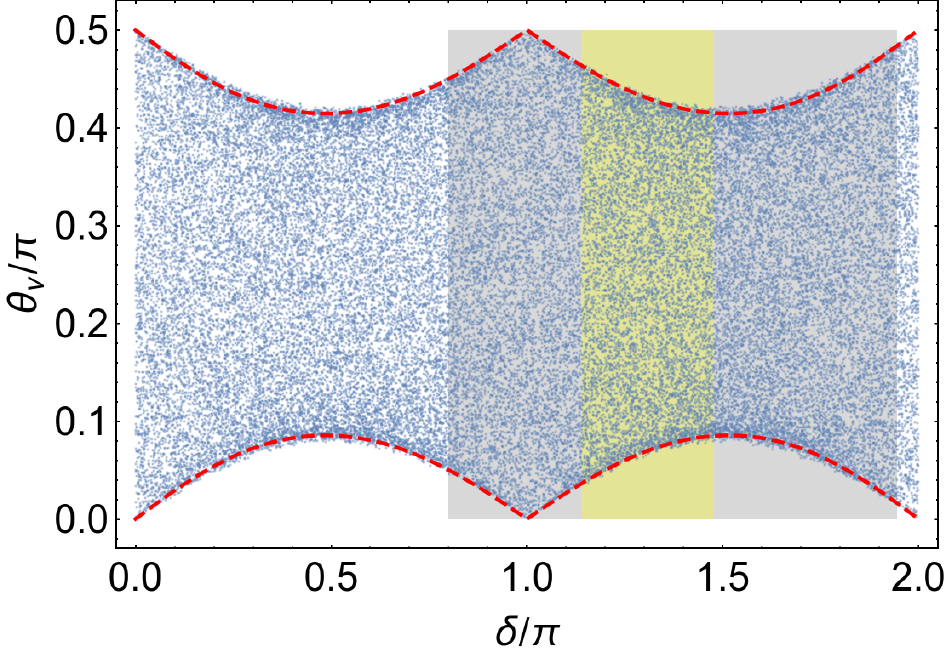}
\includegraphics[scale=.46]{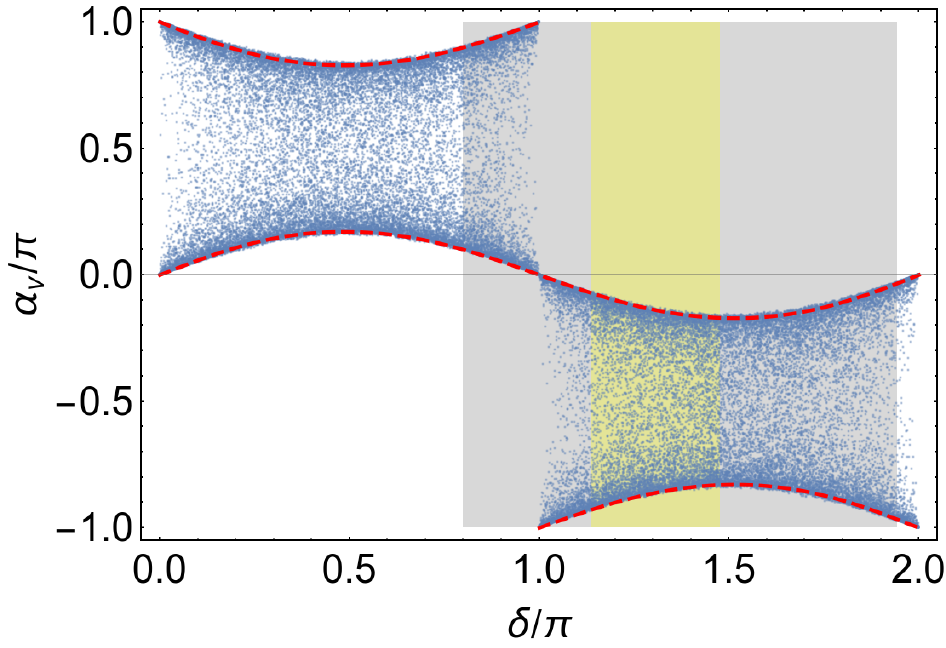}
\caption{\label{fig:Utilde:NO}
Restriction on the parameters of $\tU$ of NO parametrized as \eqref{tU:param} when 
accounting for the PMNS in \eqref{V:n=2}.
The blue points allow $3\sigma$ variation of PMNS angles while the red dashed curves uses analytical expression with best-fit angles.
}
\end{figure}

In Fig.\,\ref{fig:Utilde:NO} we show the allowed regions for $\theta_\nu$ and $\alpha_\nu$ for NO.
We clearly see that some regions are not allowed.
In particular, $\tU=\id_2$ is not possible for CP violating values of $\delta$.
More specifically, both $\theta_\nu$ and $\alpha_\nu$ cannot be zero for CP violating values of $\delta$.
For CP conserving values of $\delta$,
the parameters $\theta_\nu$ and $\alpha_\nu$ are not restricted individually.
The blue points are scattered as we vary $\theta_3'$ freely in \eqref{Utilde:OPO} and the mixing angles of the PMNS within $3\sigma$ of the global fit \cite{nufit}.
The red dashed lines are analytical expressions shown in appendix \ref{ap:n=2:R.U}, drawn using best-fit PMNS angles.
Figure\;\ref{fig:Utilde:NO} is the analogous plot for IO.
In the appendix \ref{ap:n=2:R.U} we also show that $\alpha_\nu=\pm\pi/2$ at the extrema of $\theta_\nu$ (red dashed curve) and, conversely, $\theta_\nu=\pi/4$ at the extrema of $\alpha_\nu$ (red dashed curve).
\begin{figure}[h]
\includegraphics[scale=.45]{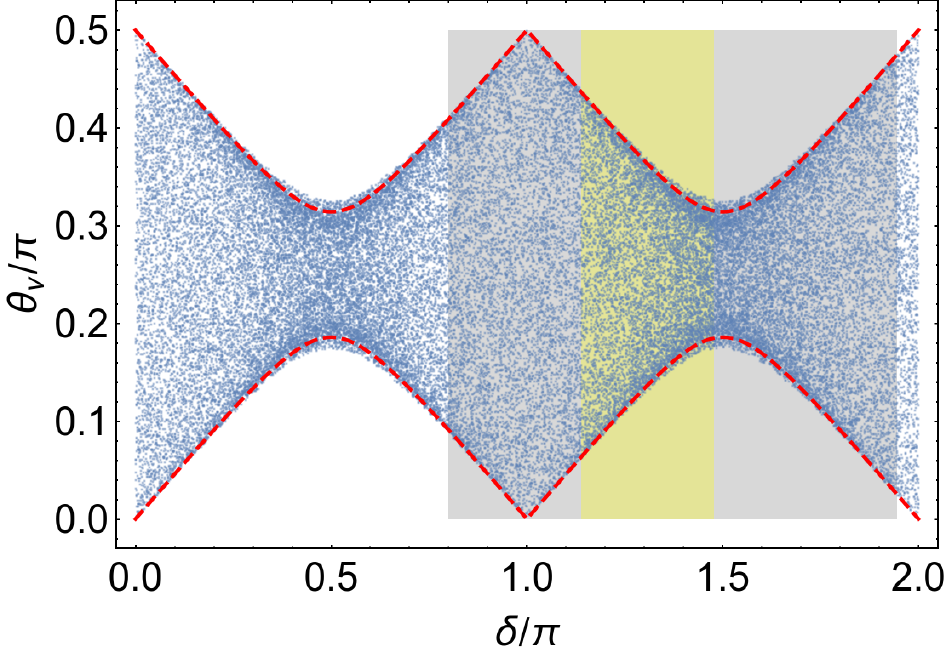}
\includegraphics[scale=.46]{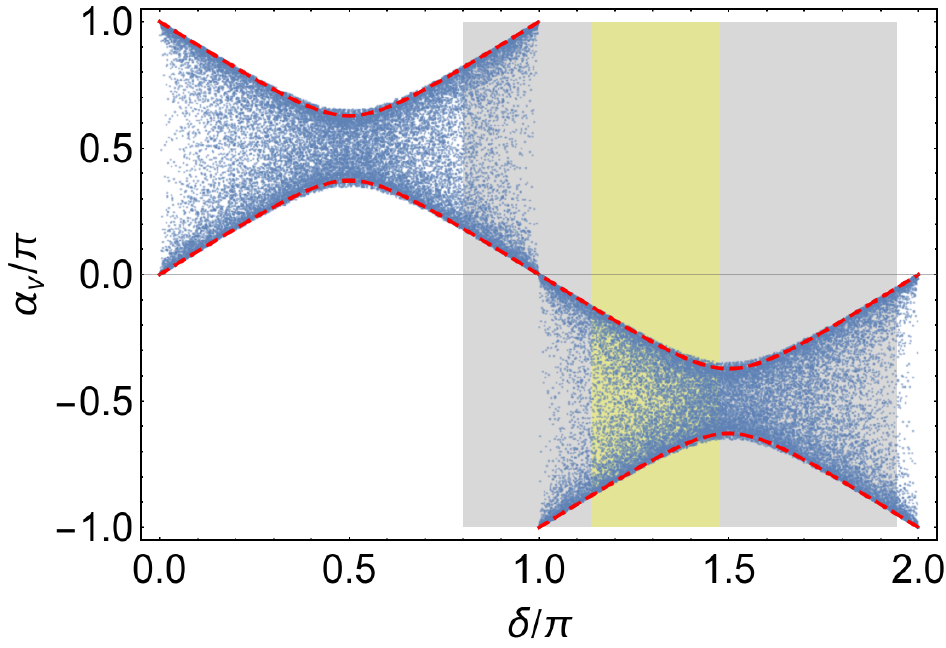}
\caption{\label{fig:Utilde:IO}
Restriction on the parameters of $\tU$ of IO parametrized as \eqref{tU:param} when 
accounting for the PMNS in \eqref{V:n=2}, adapted to  IO.
}
\end{figure}

Now we can comment on what happens in the generic CPV case. 
The real orthogonal matrix $O_e$ in \eqref{V:n=2} should be replaced by a unitary $U_e$ and its parametrization should contain one CKM-like phase $\gamma$.
We can modify the parametrization \eqref{param:n=2:Oe} to
\eqali{
\label{param:n=2:Ue}
U_e^\dag\Big|_{\rm NO}&=R_{23}(-\theta_1)R_{12}(-\theta_2)
\diag(1,e^{-i\gamma/2},e^{i\gamma/2})R_{23}(-\theta_3')\,,
\cr
U_e^\dag\Big|_{\rm IO}&=R_{13}(\theta_1)R_{23}(\theta_2)
\diag(e^{-i\gamma/2},e^{i\gamma/2},1)
R_{12}(-\theta_3')\,.
}

We have checked that the forbidden regions of Figs.\,\ref{fig:Utilde:NO} and \ref{fig:Utilde:IO} are filled when we turn on $\gamma\in [0,2\pi]$.
So indeed there are no restrictions on $\tU$ in the generic CPV case.

\subsection{Constraints on mass degeneracy of RHNs}

In the soft CPV case, we will show here that due to the restrictions on $\tU$ that enters in $M_R$ in \eqref{n=2:MR:Utilde}, mass degeneracy of $N_1,N_2$ is not possible for certain values of CP phases $\delta$ and $\beta$ of the PMNS.

For comparison, in the generic CPV case, the form of $M_R$ in \eqref{n=2:MR:Utilde} is still valid in the diagonal-$\lambda$ basis. The difference is that $\tU$ is unconstrained.
Therefore, if we choose $\tU$ diagonal in \eqref{n=2:MR:Utilde}, $M_R$ will be also diagonal so that we can easily adjust the values of the diagonal of $\lambda$ to make the masses degenerate.
Diagonal $\tU$ means that $\theta_\nu=0$
in \eqref{tU:param} ($\theta_\nu=\pi/2$ is analogous). 
As shown in Fig.\,\ref{fig:Utilde:NO}, 
for the soft CPV case, $\theta_\nu=0$ cannot be reached for CP violating values of $\delta$.

To analyze the constraints on mass degeneracy of the heavy neutrinos, we take $M_R$ in the normalized form \eqref{MR:norm} and define
\eq{
H_R\equiv\frac{M_RM_R^\dag}{M_0^2}\,.
}
Then take the characteristic polynomial
\eq{
\det\bigg(H_R-\lambda \id_2\bigg)
=\lambda^2-\gamma_1\lambda+1\,.
}
The unity comes from $\det(H_R)=1$ and the only variable coefficient is the trace
\eq{
\label{def:gamma1}
\gamma_1=\tr(H_R)\,.
}
The roots of the characteristic polynomial, which are the eigenvalues of $H_R$, are
\eq{
\label{lambda:pm}
\lambda_{\pm}=\frac{\gamma_1}{2}\pm\sqrt{\left(\frac{\gamma_1}{2}\right)^2-1}\,.
}
Since $H_R$ is hermitean and positive definite, it is guaranteed that
\eq{
\frac{\gamma_1}{2}\ge 1\,.
}
For equality, $\lambda_{\pm}=1$ and the masses are degenerate.
Given the normalization factor $M_0$ in \eqref{def:M0}, we conclude that
\eq{
\lambda_+=\frac{M_2^2}{M_0^2}=\frac{M_2}{M_1}\,,
}
is the mass ratio of the heavier to the lighter mass of $M_R$.
Note that $\lambda_-=1/\lambda_+$.

\begin{figure}[h]
\includegraphics[scale=.56
]{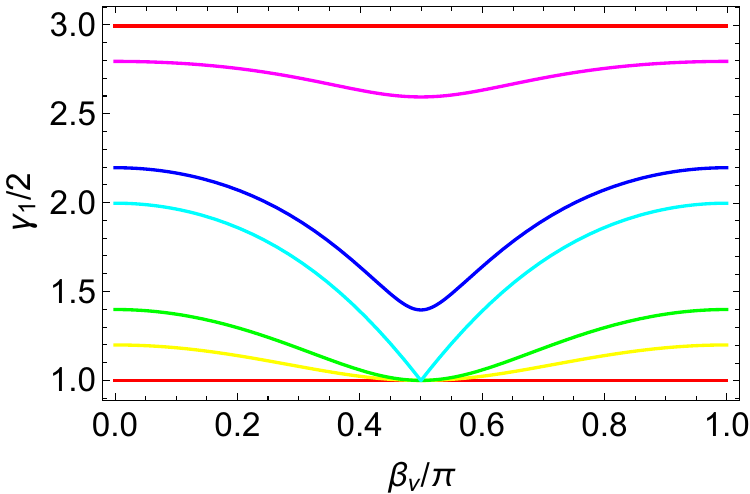}
\includegraphics[scale=.55]{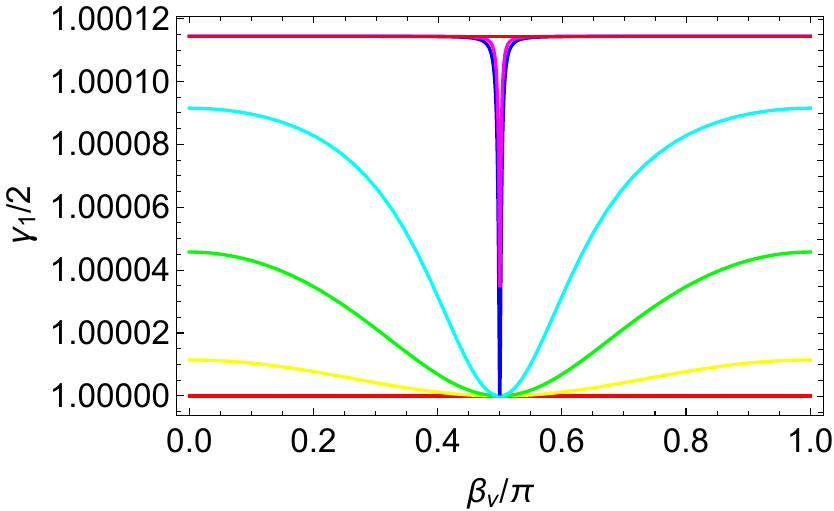}
\caption{\label{fig:trace-beta}
Half the trace \eqref{def:gamma1} as a function of the Majorana phase $\beta_\nu$ in 
\eqref{tU:param}.
Left (NO): from bottom to top, curves with $\sin^2(2\theta_\nu)=0,0.1,0.2,0.5,0.6,0.9,1$.
Right (IO): from bottom to top, curves with $\sin^2(2\theta_\nu)=0,0.1,0.4,0.8,0.9999,0.99996,1$.
}
\end{figure}

Taking the explicit form for $M_R/M_0$ in 
\eqref{MR:norm} and minimizing $\gamma_1$ with respect to $\kappa$, we obtain
\eq{
\label{gamma1:min,k}
\min_\kappa\big(\frac{\gamma_1}{2}\big)=|1+z^2|+|z|^2\,,
}
where 
\eq{
z\equiv \sin(2\theta_\nu)\ums{2}
\Big(\rho e^{i\beta_\nu}-\frac{1}{\rho e^{i\beta_\nu}}\Big)\,.
}
Note that $\gamma_1$ does not depend on $\alpha_\nu$ of $\tU$ in \eqref{tU:param} even without minimization.
In Fig.\,\ref{fig:trace-beta} we plot \eqref{gamma1:min,k} for various values of $\sin(2\theta_\nu)$ for both NO and IO using best-fit values for $\rho_{\rm NO}$ and $\rho_{\rm IO}$.
We clearly see that the minimum occurs at $\beta_\nu=\pi/2$ and that this minimum is unity up to a critical value of $\sin(2\theta_\nu)$.
This critical value is $\sin^2(2\theta_\nu)|_{\rm crit}= 0.5$ for NO and $\sin^2(2\theta_\nu)|_{\rm crit}= 0.999943$ for IO.
See appendix~\ref{ap:min:gamma1} for analytical details.

\begin{figure}[h]
\includegraphics[scale=.55]{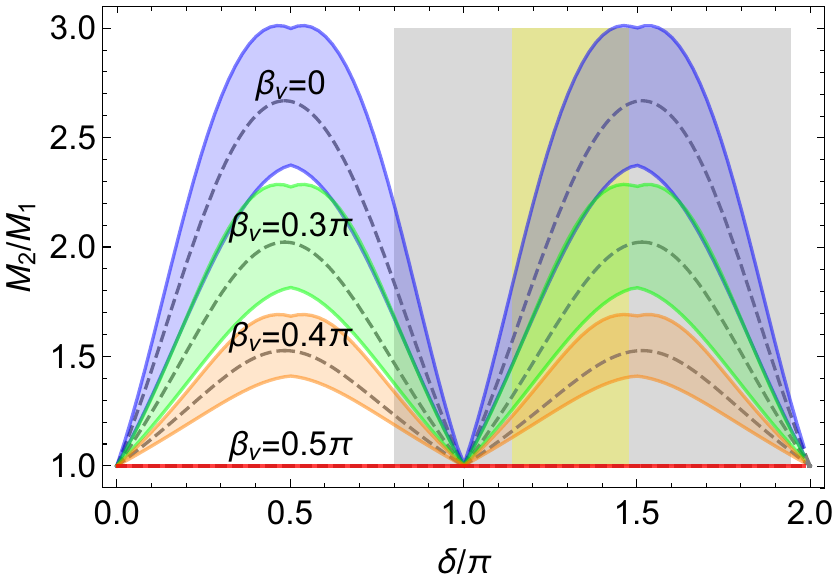}
\includegraphics[scale=.56]{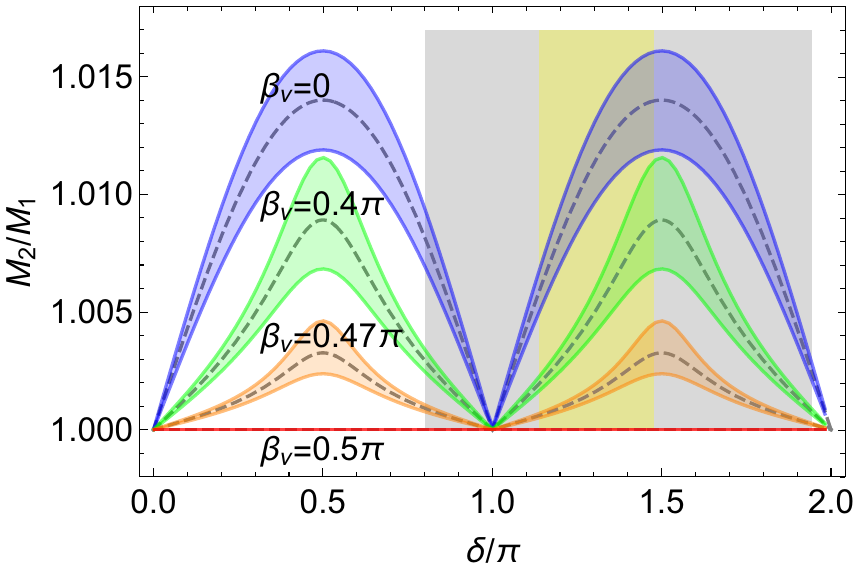}
\caption{\label{fig:min.mass.ratio}
Minimal mass ratio $M_2/M_1$ of the RHNs as a function of the Dirac phase $\delta$ for various values of the Majorana phase $\beta_\nu$ in \eqref{tU:param}.
Left: NO. Right: IO.
The bands denote the variation within $3\sigma$ of the PMNS mixing angles and mass ratios while the dashed curves are for the best-fit\,\cite{nufit}. The gray (yellow) region denotes the current $3\sigma$ ($1\sigma$) region for $\delta$.
}
\end{figure}

From figures~\ref{fig:Utilde:NO} and \ref{fig:Utilde:IO}, one can check that it is always possible to find $\sin(2\theta_\nu)$ below the critical value for any $\delta$.
So there is no absolute minimal mass ratio for $M_2/M_1$.
But, except for $\beta_\nu=\pi/2$, there is a minimal mass ratio for $M_2/M_1$ for CP violating values of $\delta$.
This result is shown in Fig.\,\ref{fig:min.mass.ratio}.
We can see the minimum mass ratio is larger for $\beta_\nu=0$ and $\delta=\pm\pi/2$.
For these values and best-fit parameters, we have
\eq{
\frac{M_2}{M_1}\Big|_{\rm NO}\ge 2.67\,,\qquad
\frac{M_2}{M_1}\Big|_{\rm IO}\ge 1.014\,
\qquad (\beta_\nu=0,\delta=\pm \pi/2).
}
For NO, the minimum above occurs at $\sin(2\theta_\nu)_{\rm min}=0.511$ and $\gamma_1/2|_{\rm min}=1.52$.
For IO, it occurs at $\sin(2\theta_\nu)_{\rm min}=0.92$ and $\gamma_1/2|_{\rm min}=1.0001$.
The minimal mass ratio for other values of $\beta_\nu$ can be seen in the figure as well as its variation due to the $3\sigma$ variation of the parameters.
For $\beta_\nu=\pi/2$ mass degeneracy is always possible.
Indeed, we show in appendix~\ref{ap:betanu=pi/2} that mass degeneracy requires $\beta_\nu=\pi/2$.

\begin{figure}[h]
\includegraphics[scale=.55]{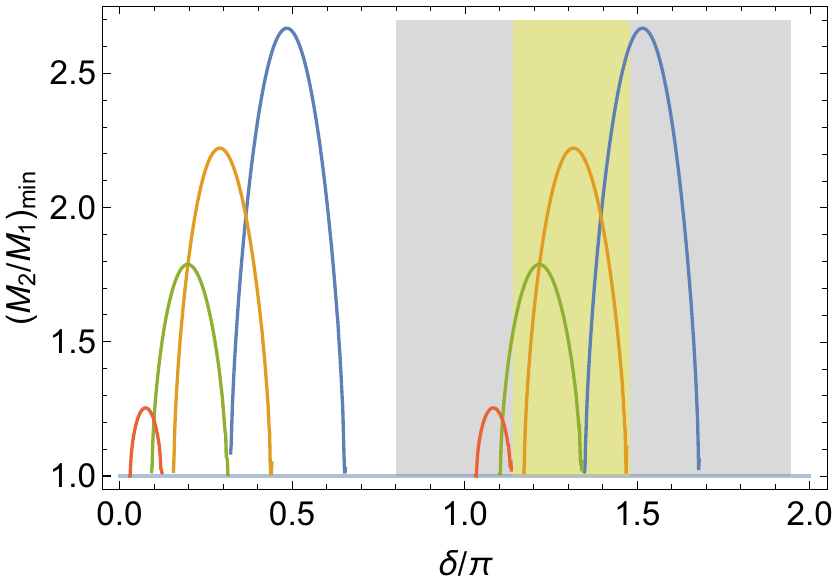}
\caption{\label{fig:min.mass.ratio:beta}
Minimal mass ratio $M_2/M_1$ of the RHNs as a funtion of the Dirac phase $\delta$ for various values of the single Majorana phase $\beta=x\times\pi/2$ with $x=0.9, 0.7, 0.5, 0$ (respectively in red, green, orange and blue) of the PMNS in \eqref{pmns:n=2} and \eqref{Ueff:n=2}.
Normal ordering.
The angles of the PMNS are of the best-fit\,\cite{nufit} while the gray (yellow) region denotes the current $3\sigma$ ($1\sigma$) region for $\delta$.
}
\end{figure}

As the phase $\beta_\nu$ is not a directly measurable quantity, it is much more interesting to show the minimum mass ratio in terms of potentially measurable quantities.
So we show in Fig.\,\ref{fig:min.mass.ratio:beta}, the minimal mass ratio for NO as a function of $\delta$, but now for various fixed values of the Majorana phase $\beta$, distinct of $\beta_\nu$, that can be potentially measured through neutrinoless double beta decay; see Fig.\,\ref{fig:beta-mbetabeta}.
Our parametrization for $\beta$ is given in \eqref{param:NO-IO}.
We clearly see below the hills the regions where mass degeneracy of the heavy neutrinos is forbidden.
Although very similar, the hills are not exactly symmetric with the shift $\delta\to \delta+\pi$.

The curves are obtained numerically by further minimizing $\gamma_1/2$ in \eqref{gamma1:min,k} as a function of the variable $\theta_3'$.
This functional dependence appears because $\alpha_\nu$ and $\theta_\nu$ in the lefthand side of \eqref{Utilde:OPO} are functions of the righthand side.
Note that because $\beta$ is fixed, these curves does not correspond to the values of $\theta_3'$ and $\beta_\nu=\pi/2$ that minimize $\gamma_1$ for free $\beta$.
There is a compromise between $\beta_\nu$ and $\theta_\nu$.
We should state that for $\beta=\pi/2$, mass degeneracy is possible while for $\beta>\pi/2$, the behavior is similar to the case of $\beta\to \pi/2-\beta$.
The curves are obtained using best-fit values of neutrino parameters but we have checked that with $3\sigma$ variation, the height of the bumps may decrease and the width decreases in the upper end.

As is clear from Fig.\,\ref{fig:min.mass.ratio}, for IO, the masses of the heavy netrinos are restricted to be different at most by around a percent. This is related to the value of $\rho_{\rm IO}$ which is very close to unity.
For this reason, a plot analogous to Fig.\,\ref{fig:min.mass.ratio:beta} for IO would show very narrow and short forbidden regions.
Therefore, we have chosen to list in table \ref{tab:min.mass.ratio:IO} only the position and the height of these forbidden regions. 
\begin{table}
\[
\begin{array}{|c|c|c|c|c|}
\hline
2\beta/\pi & \text{interval 1 $(\delta)$} & M_2/M_1-1 & \text{interval 2 $(\delta)$} & M_2/M_1-1 \\
\hline
  0  & [1.6815, 1.696] & 0.0152408 & [4.5872, 4.6017] & 0.0152410
  \cr
\hline
  0.5  & [2.1573, 2.1739] & 0.0152411 & [4.9189, 4.9349] & 0.0152411
  \cr  
\hline
  0.7  & [2.5057, 2.518] & 0.0152411 & [5.1176, 5.1371] & 0.0152411
  \cr  
\hline
  0.9  & [2.9253, 2.9295] & 0.0152398 & [5.5359, 5.5641] & 0.0152411
  \cr  
\hline
  0.99  & [3.11993, 3.12035] & 0.0152393 & [6.1517, 6.1651] & 0.0152409
    \cr
\hline
\end{array}
\]
\caption{\label{tab:min.mass.ratio:IO}
Excluded regions for IO of minimal mass ratio $M_2/M_1$ of the RHNs as a funtion of the Dirac phase $\delta$ for various values of the single Majorana phase $\beta$ of the PMNS in \eqref{pmns:n=2} and \eqref{Ueff:n=2}.
Similarly to Fig.\,\ref{fig:min.mass.ratio:beta}, 
the intervals denote the regions where $M_2>M_1$ and $M_2/M_1-1$ denotes the peak value. 
The angles of the PMNS are of the best-fit\,\cite{nufit} while the gray (yellow) region denotes the current $3\sigma$ ($1\sigma$) region for $\delta$.
}
\end{table}

\begin{figure}[h]
\includegraphics[scale=.47]{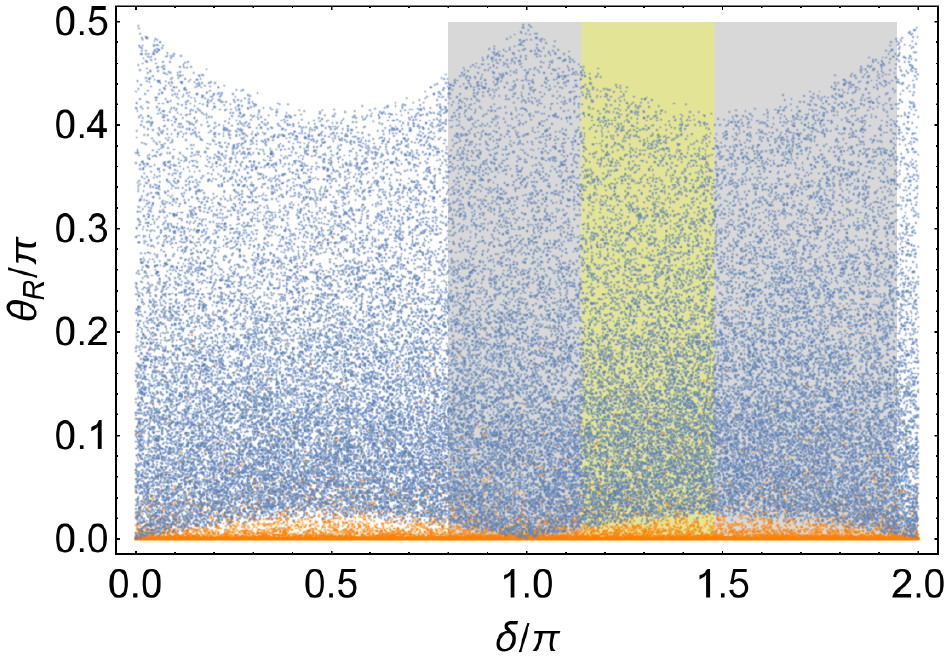}
\includegraphics[scale=.47]{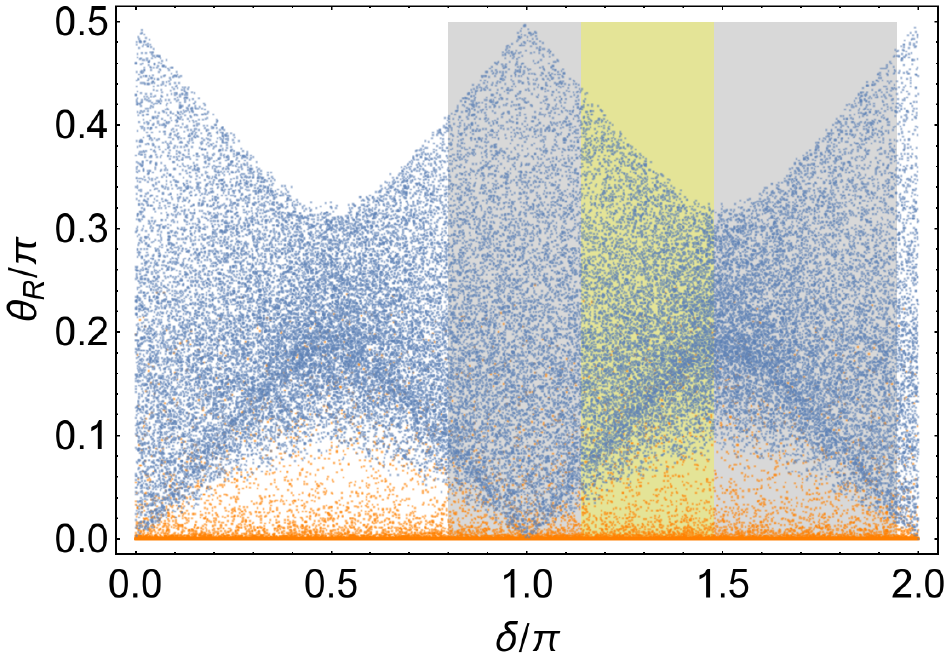}
\caption{\label{fig:UR}
Scatter plot for $\theta_R$, parametrized in \eqref{UR:param}, as a function of the Dirac CP phase $\delta$.
For NO (left), the points in blue (orange) denote $\kappa\in [0.5,1]$ ($\kappa\in [0.01,0.5]$), where $\kappa$ is the variable in \eqref{MR:norm}.
For IO (right), the points in blue (orange) denote $\kappa\in [0.995,1]$ ($\kappa\in [0.01,0.995]$).
The squared mass differences and angles of the PMNS are varied within $3\sigma$.
Thee gray (yellow) region denotes the current $3\sigma$ ($1\sigma$) region for $\delta$.
}
\end{figure}

At last, we show the restrictions on $U_R$ that diagonalizes $M_R$ as
\eq{
U_R^\dag M_RU_R^*=\diag(M_1,M_2)\,,
}
where $M_1<M_2$ is chosen.
Through \eqref{MR:norm}, $U_R$ will be a function of $\tU$ and restrictions on the latter will lead to restrictions on the former.
By using the same parametrization as $\tU$,
\eq{
\label{UR:param}
U_R=W(\alpha_R,\theta_R,\beta_R)\,,
}
we can see in Fig.\,\ref{fig:UR} that there are forbidden regions in the plane $\theta_R$ and $\delta$.
Noting that in \eqref{MR:norm} we can restrict ourselves to either $\kappa\ge 1$ or $\kappa\le 1$ as a generic $\tU$ allow the exchange $\kappa\longleftrightarrow \kappa^{-1}$.
We have chosen $\kappa\le 1$. If we had used $\kappa \ge 1$, we would have obtained forbidden angles $\theta_R$ close to zero.
As before, if we turn on the additional phase $\gamma$ in \eqref{param:n=2:Ue} by replacing $O_e$ by $U_e$, the forbidden regions are eliminated and no restriction remains.

%%%%%%%%%%%%%%%%%%%%%%%%%%%%%%%%%%%%%%%%%%%%%%%%%
\section{Conclusions}
\label{sec:concl}

We have studied the simple case where the seesaw mechanism is simultaneously responsible for the light neutrino masses and the leptonic CP violation. The only source for CP breaking is soft and resides in the Majorana mass term of the heavy righthanded neutrinos.
We denoted this case as the soft CP violation case in contrast to the generic CP violation case.
This soft breaking can be easily attributed to scalar field vevs. 

For the realistic case of two or more RHNs, the soft CP violating Lagrangian depends on one less parameter compared to the generic CP violation case.
This reduction of one less parameter is common to other scenarios where CP violation in the SM are generated spontaneously\,\cite{nb-vlq,nb-vll}.
For the seesaw case, this parameter reduction leads to subtle restrictions of the seesaw parameter space once leptonic CP violation is assumed and the current knowledge of neutrino masses and mixing are taken into account.
The restrictions act on the high energy sector and the low energy parameters are not restricted in any way when seen in isolation. 
For the minimal seesaw case, we analyzed the restricted parameters in detail, most notably, 
there are correlations between the low energy CP phases of the PMNS with the mass splitting of RHNs. For certain values, mass degeneracy is forbidden and resonant leptogenesis is severely restricted unless freeze-in lepetogenesis effects are taken into account\,\cite{shaposhnikov,drewes}.

%%%%%%%%%%%%%%%%%%%%%%%%%%%%%%%%%%%%%%%%%%%
\acknowledgments

The authors thank Gustavo Branco and Gui Rebelo for useful discussions in the beginning of this project.
C.C.N.\ acknowledges partial support by Brazilian Fapesp, grant 2014/19164-6, and CNPq, grant 312866/2022-4.

%%%%%%%%%%%%%%%%%%%%%%%%%%%%%%%%%%%%%%%%%%%%%%%%%
\appendix
%%%%%%%%%%%%%%%%%%%%%%%%%%%%%%%%%%%%%%%%%%%%%%%%%%%%%%%%%%
\section{Orthogonal decomposition}
\label{ap:ortho.decomp}

One can show that a $n\times n$ unitary matrix $U$ can be always decomposed into
\eq{
\label{ortho.param}
U=\cO_2\hat{U}\cO_1^\tp\,,
}
where $\cO_i$ are real orthogonal matrices and $\hat{U}$ is a diagonal matrix of phases (they are not the eigenvalues of $U$).
The number of parameters are $n\times(n-1)/2$ in each $\cO_i$ and $n$ phases in $\hat{U}$, so $n^2$ in total as it should be. One phase in $\hU$ can be removed if global rephasing is allowed.

The proof is simple.
We first check that
\eq{
4i[\re(U^\tp U),\im(U^\tp U)]
=[U^\tp U+(U^\tp U)^*,U^\tp U-(U^\tp U)^*]
=-2[U^\tp U,(U^\tp U)^*]=0\,.
}
Then the real and imaginary parts of the unitary symmetric matrix $U^\tp U$ can be diagonalized simultaneously by a real orthogonal matrix:
\eq{
\cO_1^\tp U^\tp U\cO_1=\diag(e^{i2\lambda_j})=\hat{U}^2\,.
}
The diagonalizing matrix is $\cO_1$ and the diagonal matrix is $\hat{U}^2$.
Then $U\cO_1$ coincides with $\hat{U}$ except for a rotational freedom from the left which we denote as $\cO_2$ resulting in
\eq{
\label{U.O1}
U\cO_1=\cO_2\hat{U}\,.
}
As the eigenvectors in $\cO_1$ has no definite global sign, we can conventionally choose the phases in $\hU$ to lie in $[0,\pi)$.

As a special case, let us assume $U$ admits rephasing transformations from the left.
Then we can always take, e.g., the first column real and non-negative. In this case $U^\tp U$ is block diagonal
\eq{
\label{UU^t:R}
U^\tp U=\mtrx{1&0&\dots\cr 0&*&*\cr\vdots&*&*}\,,
}
because $(U^\tp U)_{11}=(U^\dag U)_{11}=1$ already. In this case $\cO_1$ only needs to rotate in $n-1$ dimensions and depends only on $(n-1)(n-2)/2$ angles. It is also clear from \eqref{UU^t:R} that $\hat{U}$ has $\hat{U}_{11}=1$ and $\hat{U}$ depends only on $n-1$ phases.
As $\cO_2$ contains the usual $n(n-1)/2$ angles, the total amounts to $n(n-1)$ parameteres, a reduction in $n$ parameters corresponding to the rephasing of $n$ phases.

For $n=3$, this result has implications on the parametrization of the PMNS matrix $V$ which is physically equivalent under rephasing from the left.
The parametrization \eqref{ortho.param} allows us to parametrize the PMNS matrix with \emph{only two phases}.
Specifically in the rephasing convention \eqref{UU^t:R} we can write
\eq{
% \label{2phases}
V=\cO_2\diag(1,e^{i\alpha/2},e^{i\beta/2})R_{23}(\theta_{23})\,,
}
where $R_{23}$ is a rotation in the $(23)$ plane and $\alpha,\beta$ are confined to $[0,2\pi]$ due to sign flip freedom from the right. The angle $\theta_{23}\in [0,2\pi)$ as well. The matrix $\cO_2$ is a generic $SO(3)$ matrix that can be parametrized with three Euler angles. Hence $V$ in \eqref{2phases} is parametrized with 4 angles and 2 phases.
Therefore, one phase (CP odd) is traded by one angle (CP even) if compared to the standard parametrization.

\section{Orthogonal decomposition: rephasing from both sides}
\label{ap:orthog.2}

The analysis is slightly more complicated if the unitary matrix $U$ is allowed rephasing transformations from both sides. Here we analyze this case specifically for $n=3$.
The result is that the orthogonal decomposition \eqref{ortho.param} is possible with \emph{only one phase}.
This suggests a different parametrization for the CKM matrix but the number of phases is unchanged.

Let us denote the columns of $U$ as
\eq{
U=\left(
\begin{array}{c|c|c}
u_1 & u_2 & u_3 
\end{array}
\right)\,,
}
and analogously for $U^\dag$
\eq{
U^\dag=\left(
\begin{array}{c|c|c}
v_1 & v_2 & v_3 
\end{array}
\right)
=\left(
\begin{array}{c}
u_1^\dag\cr\hline
u_2^\dag\cr\hline
u_3^\dag
\end{array}
\right)
\,.
}

If $e_i$ are the canonical column vectors, we can write
\begin{align}
\label{Uei}
Ue_i=u_i\,, &\quad
Uv_i=e_i\,, \cr
U^\dag e_i=v_i\,, &\quad
U^\dag u_i=e_i\,, 
\end{align}
for $i=1,2,3$.

By choosing $u_1$ real by rephasing from left, we obtain
\eq{
U^\tp u_1=e_1\implies
U^\tp Ue_1=e_1\,,
}
and $U^\tp U$ is block diagonal as in \eqref{UU^t:R}.
If additionally $v_1=U^\dag e_1$ is real by rephasing from the right, then also $v_1=U^\tp e_1$ and
\eq{
U^\tp Uv_1=U^\tp e_1=v_1\,.
}
Assuming $v_1\neq e_1$, which is equivalent to $U$ non-block-diagonal, we conclude that the eigenvalue $\lambda=1$ of $U^\tp U$ has multiplicity at least two. Discarding real $U$, the multiplicity should be exactly two.

Now construct $\cO_1$ with columns $e_1,v_1',v_1''$, where $(e_1,v_1')$ is the orthogonal basis for the space spanned by $(e_1,v_1)$ while $v_1''$ is the orthogonal direction ($\cO_1$ is block diagonal).
It is clear that 
\eq{
\label{Uhat:both}
\cO_1^\tp U^\tp U\cO_1=\diag(1,1,e^{i2\beta})=\hat{U}^2\,,
}
where incidentally $e^{i\beta}=\det U$.
Equation \eqref{U.O1} tells us that the first column of $\cO_2$ is $u_1$, i.e., the same as $U$'s.
Since $Uv_1=e_1$, the second column of $\cO_2$ is the piece of $e_1$ orthogonal to $u_1$.
So we can obtain $\cO_1$ and $\cO_2$ from orthogonalization of
\eq{
\label{O1.O2:ortho}
\cO_1\leftarrow (e_1,v_1,e_1\times v_1)\,,\quad
\cO_2\leftarrow (u_1,e_1,u_1\times e_1)\,.
}
In this case $\cO_1$ is block diagonal and $\cO_2$ has vanishing $(13)$ entry.
We can reverse the roles of $\cO_1$ and $\cO_2$ making the latter block diagonal. Other choices of real columns and rows lead to similar forms.

The parameter counting goes as follows: $\hU$ contains one phase, $\cO_1$ depends on one angle to define $v_1$ in the subspace orthogonal to $e_1$ and $\cO_2$ depends on two angles to define $u_1$. The total amounts to the correct number of 4 parameters

For $n>3$, the construction of the first two columns in $\cO_1$ and $\cO_2$ continues to be as in \eqref{O1.O2:ortho}. For $\cO_1$, the first two columns depend on the pair $(e_1,v_1)$ which requires $n-2$ angles to define the part of $v_1$ orthgonal to $e_1$. The rest of the $n-2$ columns span an $n-2$ dimensional space which can be rotated by the $O(n-2)$ group with $(n-2)(n-3)/2$ angles. For $\cO_2$, the reasoning is the same, except that the pair $(u_1,e_1)$ needs $n-1$ angles to define $u_1$.
As the number of phases in \eqref{Uhat:both} becomes $n-2$, the total number of parameters is $2\times(n-2)+(n-1)+2\times (n-2)(n-3)/2=(n-1)^2$, matching the number of parameters for a CKM-like mixing matrix with $n$ families\,\cite{branco:book}.
The number of phases here, however, differ from the standard counting (parametrization).
The orthogonal decomposition uses $n-2$ phases. The standard parametrization requires $(n-1)(n-2)/2$.
For $n=3$, one phase is required in both cases (the KM counting) but for larger $n$ the scaling is linear rather than quadratic and much less phases appear.

\section{Extrema of parameters for $n_R=2$}
\label{ap:n=2:R.U}

For $n_R=2$, we are confronted with the problem of mapping the possible parameter space of the $SU(2)$ matrix 
\eq{
\label{U:AB}
\mtrx{A & -B^*\cr B& A^*}
=\mtrx{c_\theta & s_\theta \cr -s_\theta & c_\theta}\mtrx{a & -b^*\cr b& a^*}\,,
}
with fixed $a,b$ and varying $\theta$.
These matrices belong to $SU(2)$ so that $|a|^2+|b|^2=1$ and $|A|^2+|B|^2=1$.

Parametrizing
\eq{
|a|=c_{\theta'}\,,\quad
|b|=s_{\theta'}\,,\quad
ba^*=|ab|e^{i\alpha'}\,,
}
with $\theta'\in [0,\pi/2]$, we find
\eq{
2|B|^2=1-c_{2\theta}c_{2\theta'}-s_{2\theta}s_{2\theta'}c_{\alpha'}\,.
}
This quantity has extrema at
\eq{
(c_{2\theta},s_{2\theta})=\pm
\frac{1}{N}(c_{2\theta'},s_{2\theta'}c_{\alpha'})\,,
}
where $N$ is the normalization factor.
Hence, the minimal and maximal values of $|B|^2$ are respectively given by
\eq{
2|B|^2_{\stackrel{\text{\scriptsize{min}}}{\text{\scriptsize{max}}}}
=1\mp \sqrt{1-s_{2\theta'}^2s^2_{\alpha'}}\,.
}
Note that $|A|^2_{\rm min}=1-|B|^2_{\rm max}$ and analogously for the maximum.

We can check that, at the extrema,
\eq{
BA^*\big|_{\rm extr}=\frac{i}{2}s_{2\theta'}s_{\alpha'}\,,
}
so that the relative phase of $A,B$ is $\pm \pi/2$.
Therefore, for the minimum of $|B|$, the matrix in \eqref{U:AB} can be written as
\eq{
e^{\mp i\sigma_3\pi/4}
\mtrx{|A|_{\max} & -|B|_{\min} \cr |B|_{\min} & |A|_{\max}}e^{-i\sigma_3\beta/2}\,,
}
where the minus (plus) sign is for $s_{\alpha'}>0$ ($s_{\alpha'}<0$).
The phase $\beta$ depends on other parameters.
The case of maximum $|B|$ is similar.

If we want to extremize $BA^*$ with respect to $\theta$, then
\eq{
(c_{2\theta},s_{2\theta})=\pm \frac{1}{N}(-s_{2\theta'}c_{\alpha'},c_{2\theta'})\,.
}
The extrema are
\eq{
BA^*\big|_{\rm ext}=is_{2\theta'}s_{\alpha'}\mp \sqrt{1-s^2_{2\theta'}s^2_{\alpha'}}\,.
}
At these extrema,
\eq{
|A|^2=|B|^2=\frac{1}{2}\,.
}

%%%%%%%%%%%%%%%%%%%%%%%%%%%%%%%%%%%%%%%%%%%%%%%%%
\section{Minimizing $\gamma_1$}
\label{ap:min:gamma1}

From \eqref{MR:norm}, the trace $\gamma_1$ is
\eq{
\gamma_1=\Tr[H_R]
=\Tr[R_\rho\diag(\kappa^2,\kappa^{-2})R_\rho^*\diag(\kappa^2,\kappa^{-2})]\,,
}
where
\eq{
R_\rho=\tU \diag(\rho,\rho^{-1})\tU^\tp
=\mtrx{A &B\cr B & D}
\,.
}

The minimum
\eq{
\min_\kappa(\gamma_1)=2(|AD|+|B|^2)\,,
}
occurs at
\eq{
\kappa^4=\frac{|D|}{|A|}\,.
}
Use of the parametrization \eqref{tU:param} of $\tU$ leads to \eqref{gamma1:min,k}.
It is also clear that $\gamma_1$ does not depend on $\alpha_\nu$.

Further maximization of \eqref{gamma1:min,k} in $\beta_\nu$ yields
\eq{
\max_{\beta_\nu}\min_\kappa(\gamma_1/2)=
1+2z^2_{\min}\,,
}
where $z_{\min}=\sin(2\theta_\nu)(\rho-1/\rho)/2$ for $\beta_\nu=0,\pi$.
Further minimization of \eqref{gamma1:min,k} in $\beta_\nu$ yields
\eq{
\min_{\beta_\nu}\min_\kappa(\gamma_1/2)=
|1-z_{\max}^2|+z^2_{\max}\,,
}
where $z_{\max}=\sin(2\theta_\nu)(\rho+1/\rho)/2$ for $\beta_\nu=\pm\pi/2$.
The previous value is unity as long as $z_{\rm max}\le 1$.
This means that $\min_\kappa(\gamma_1/2)=1$ for $\beta_\nu=\pm\pi/2$ as long as 
\eq{
\sin(2\theta_\nu)\le 2(\rho+1/\rho)^{-1}\,,
}
where the value on the righthandside is less than unity.
For $\sin(2\theta_\nu)$ above this critical value, $\gamma_1/2> 1$ is guaranteed and degeneracy of $N_1,N_2$ is not possible; see Fig.\,\ref{fig:trace-beta}.
For best-fit values of $\rho$, the critical value above is 0.7076 for NO and 0.999971 for IO.
The mass ratio is obtained from $\gamma_1/2$ from \eqref{lambda:pm}.

%%%%%%%%%%%%%%%%%%%%%%%%%%%%%%%%%%%%%%%%%%%%%%%%%
\section{Direct proof of $\beta_\nu=\pi/2$ for degenerate $N_1,N_2$.}
\label{ap:betanu=pi/2}

For $n_R=2$, we have seen in Fig.\,\ref{fig:min.mass.ratio} that degenerate RHNs (for any $\delta$) required $\beta_\nu=\pi/2$. 
Here we prove this fact directly.

For $M_1=M_2$, the normalized mass matrix is
\eq{
\frac{M_R}{M_0}=U_RU_R^\tp\,,
}
where the diagonalization matrix $U_R$ is defined up to the rotation freedom $U_R\to U_RO$.
Since $M_R/M_0$ is a unitary and symmetric matrix, we can always parametrize it as
\eq{
\frac{M_R}{M_0}=e^{-i\sigma_3\gamma/2}\mtrx{c_\theta & s_\theta \cr s_\theta & -c_\theta}e^{-i\sigma_3\gamma/2}\,,
}
already discarding a global phase.

Then, for both NO and IO, the nontrivial block matrix \eqref{n=2:Mtil} for light neutrinos becomes
\eq{
\label{tMnu:degere}
\tilde{M}_\nu=-\frac{v^2}{M_0}\tlamb\, 
    e^{+i\sigma_3\gamma/2}
    \mtrx{c_\theta & s_\theta \cr s_\theta & -c_\theta}
    e^{+i\sigma_3\gamma/2}\,
    \tlamb
\,.
}
Since $\tlamb=\diag(\lambda_1,\lambda_2)$ commutes with $e^{+i\sigma_3\gamma/2}$, 
we can see that the matrix $\tU$ that diagonalizes $\tilde{M}_\nu$ is
\eq{
\tU=e^{-i\sigma_3\gamma/2}O\diag(1,i)\,,
}
where $O$ is a real orthogonal matrix.
The last phase $i$ corresponds exactly to $\beta_\nu=\pi/2$ in the parametrization \eqref{tU:param} and appears because the matrix in the middle of \eqref{tMnu:degere} involving $\theta$ and $\tlamb$ has real eigenvalues but of opposite signs.

%%%%%%%%%%%%%%%%%%%%%%%%%%%%%%%%%%%%%%%%%%%

%%%%%%%%%%%%%%%%%%%%%%%%%%%%%%%%%%%%%%%%%%%%%%%%%
\end{document}